\documentclass[10pt,twocolumn,superscriptaddress,floatfix,nobalancelastpage,amsmath,longbibliography]{revtex4-1}

\usepackage{array,longtable}
\newcolumntype{L}{>{\tiny $}p{0.33\columnwidth}<{$}}
\newcolumntype{M}{>{\scriptsize $}p{0.33\columnwidth}<{$}}
\newcolumntype{N}{>{\scriptsize $}p{0.43\columnwidth}<{$}}
\usepackage{amsmath}
\usepackage{amssymb}
\usepackage{amsfonts}
\usepackage{graphicx}
\usepackage{hyperref}
\usepackage{mathbbol}
\usepackage[vcentermath]{youngtab}
\usepackage{xcolor}

\usepackage{multirow}
\usepackage{float}
\usepackage{graphics}
\usepackage{physics}
 \usepackage{amssymb}

\allowdisplaybreaks[1]

\newif\ifhyper
\hypertrue
\ifhyper
\hypersetup{
   citecolor = {green},
   colorlinks = {true}, % false
   urlcolor = {blue} % magenta
}

\usepackage{array,longtable}
\usepackage{multirow}
\newcolumntype{C}{>{$}c<{$}}    % math-mode version of "c" column type

\def \sym{\; \color{green} \checkmark \color{black} \;}
\def \nsym{\; \color{red} \times \color{black} \;}

\begin{document}

%%%%%%%%%%%%%%%%%%%%%%%%%%%%%%%%%%%%%%%%%%%%%%%%%%%%%%

\title{
Quantum spin liquid phases in the bilinear-biquadratic two-SU(4)-fermion Hamiltonian on the square lattice
}
%%%%%%%%%%%%%%%%%%%%%%%%%%%%%%%%%%%%%%%%%%%%%%%%%%%%%%

%
\author{Olivier Gauth\'e}
\affiliation{Laboratoire de Physique Th\'eorique, Universit\'e de Toulouse, CNRS, UPS, France}
\affiliation{Institute of Theoretical Physics, \'Ecole Polytechnique F\'ed\'erale de Lausanne (EPFL), CH-1015 Lausanne, Switzerland}
\author{Sylvain Capponi}
\affiliation{Laboratoire de Physique Th\'eorique, Universit\'e de Toulouse, CNRS, UPS, France}
\author{Matthieu Mambrini}
\affiliation{Laboratoire de Physique Th\'eorique, Universit\'e de Toulouse, CNRS, UPS, France}
\author{Didier Poilblanc}
\affiliation{Laboratoire de Physique Th\'eorique, Universit\'e de Toulouse, CNRS, UPS, France}
\affiliation{Institute of Theoretical Physics, \'Ecole Polytechnique F\'ed\'erale de Lausanne (EPFL), CH-1015 Lausanne, Switzerland}

%%%%%%%%%%%%%%%%%%%%%%%%%%%%%%%%%%%%%%%%%%%%%%%%%%%%%%

\date{\today}

%%%%%%%%%%%%%%%%%%%%%%%%%%%%%%%%%%%%%%%%%%%%%%%%%%%%%%

\begin{abstract}
We consider the phase diagram of the most general SU(4)-symmetric two-site Hamiltonian for a system of two fermions per site (i.e.\ self-conjugate $\bf 6$ representation) on the square lattice. 
It is known that this model hosts magnetic phases breaking SU(4) symmetry and quantum disordered dimer-like phases breaking lattice translation symmetry. Motivated by
a previous work [O. Gauth\'e, S. Capponi and D. Poilblanc, Phys. Rev. B \textbf{99}, 241112(R) (2019)], we investigate the possibility of the existence of SU(4) quantum spin liquid phases in this model, using SU(4)-symmetric Projected Entangled Pair States (PEPS) of small bond dimensions, which can be classified according to point group and charge (C) symmetries. Among several (disconnected) families of SU(4)-symmetric PEPS, breaking or not C-symmetry, we identify critical or topological spin liquids which may be stable in some regions of the phase diagram. These results are confronted to exact diagonalization (ED) and density matrix renormalization group (DMRG) calculations. 
\end{abstract}

%%%%%%%%%%%%%%%%%%%%%%%%%%%%%%%%%%%%%%%%%%%%%%%%%%%%%%
\maketitle

%%%%%%%%%%%%%%%%%%%%%%%%%%%%%%%%%%%%%%%%%%%%%%%%%%%%%%%%%%%%%%%%%%%%%%%%%%%%%%%%%%%%%%%%%%%%%%%%%%%%%%%%%
\section{Introduction}

With the realization of ultracold gases of atoms with $N$ internal (nucleus) degrees of freedom loaded on periodic optical lattices~\cite{cazalilla_ultracold_2014,hofrichter_direct_2016}, an interest is rapidly growing for spin Hamiltonians with exact $\mathrm{SU}(N)$ symmetry.  Various lattices, $\mathrm{SU}(N)$ symmetries and $\mathrm{SU}(N)$ irreducible representations (irreps) have been studied~\cite{sutherland_model_1975, read_features_1989, read_spin-peierls_1990, corboz_simultaneous_2011, bauer_three-sublattice_2012, capponi_phases_2016}, showing plethora of novel phases, most of them spontaneously breaking lattice or $\mathrm{SU}(N)$ symmetries, like Valence Bond Crystals (VBC) or magnetic states. However, a few studies were devoted to the explicit construction of $\mathrm{SU}(N)$ quantum spin liquids (QSL) preserving both $\mathrm{SU}(N)$ and lattice symmetries~\cite{gauthe_entanglement_2017,dong_su3_2018,kurecic_gapped_2019}. Tensor networks like Projected Entangled Pair States (PEPS) are particularly well suited to the construction of QSL states.
For example, a previous work proposed critical QSL and a $\mathbb{Z}_2$ topological QSL phase for a system of two SU(4) fermions per site. 
Although VBC are ubiquitous in the study of $\mathrm{SU}(N)$-invariant models, (nonchiral) spin liquids seem to be relatively rare.
Here we revisit a SU(4)-symmetric bilinear-biquadratic Hamiltonian with two fermions in the self-conjugate $\bf 6$-irrep of SU(4) on each site~\cite{assaad_phase_2005,kim_linear_2017,kim_dimensional_2019}. 
Using exact diagonalization (ED), density matrix renormalization group (DMRG) and infinite-PEPS (iPEPS) numerical methods, we identify two different types of SU(4) spin liquids which appear to be very competitive in energy in two regions of the (one parameter) phase diagram. 

\section{Model and Hamiltonian}

We consider a square lattice where we attach a SU(4) irreducible representation \hbox{$\textbf{6}\equiv\tiny{\yng(1,1)}$} corresponding to the six antisymmetric states of two SU(4) (atomic) fermions on each site.
 We also assume a coupling between nearest-neighbor (NN) sites only. Starting from the fusion rule on two sites
\begin{equation}
\yng(1,1) \, \otimes \, \yng(1,1)\; = \; \bullet \, \oplus \, \yng(2,1,1) \, \oplus \, \yng(2,2)  \,\, ,
\label{Eq:fusion}
\end{equation}
we see that three SU(4) symmetric projectors can be defined on these two sites: $\mathcal{P}_\textbf{1}$, $\mathcal{P}_\textbf{15}$ and $\mathcal{P}_\textbf{20'}$, corresponding to the fusion outcomes characterized by the irreps $\bf 1$, $\bf 15$ and $\bf 20'$, on the right hand side of Eq.~\ref{Eq:fusion}, respectively. 
One can use the projectors $\mathcal{P}_\alpha$ as a natural basis to expand the Hamiltonian $\mathcal{H}=\sum_\alpha c_\alpha\mathcal{P}_\alpha$, $c_\alpha\in \mathbb{R}$.
The operator $\textbf{S}\cdot \textbf{S}$ on two sites writes
%\begin{equation}
$\textbf{S}\cdot \textbf{S} = -5 \mathcal{P}_\textbf{1} - \mathcal{P}_\textbf{15} + \mathcal{P}_\textbf{20'}$, 
%\end{equation}
which, as can be seen straightforwardly, is linearly independent from $(\textbf{S}\cdot \textbf{S})^2=25\mathcal{P}_\textbf{1} + \mathcal{P}_\textbf{15} + \mathcal{P}_\textbf{20'}$. As $\sum_\alpha \mathcal{P}_\alpha=\mathcal{I}_d$, 
the most general two-sites SU(4) symmetric (real) Hamiltonian can then be re-expressed as a linear combination of $\textbf{S}\cdot \textbf{S}$ and $(\textbf{S}\cdot \textbf{S})^2$ (up to a constant energy shift) and can be parametrized by a single parameter $\theta$. Following the conventions of \cite{affleck_su2n_1991}, the lattice Hamiltonian becomes
\begin{equation}
\mathcal{H}(\theta) = \cos\theta \; \sum_{\big< i j\big>}\textbf{S}_i \cdot \textbf{S}_j  +  \frac{\sin\theta}{4} \, \sum_{\big< i j\big>}(\textbf{S}_i\cdot \textbf{S}_j)^2\, ,
\label{eq:H66}
\end{equation}
where the sum is restricted to nearest-neighbor bonds $\big< ij\big>$.
In addition to the invariance w.r.t.\ the lattice symmetries and the SU(4) spin symmetry, Hamiltonian (\ref{eq:H66}) is also invariant w.r.t.\ color (or ``charge'') conjugation (C) since physical degrees of freedom correspond to a self-conjugate irrep of SU(4). 

Importantly, there are four SU(6) points, when the coefficients in front of two projectors are identical, and the fusion rules are enhanced to that of SU(6):
i) at $\theta = \pi/4$ and $\theta = -3\pi/4$, the fusion rule is enhanced to $\textbf{6} \otimes \textbf{6} = \textbf{15} \oplus \textbf{21}$ and 
ii) at $\theta = \pm \pi/2$, the fusion rule becomes $\textbf{6} \otimes \overline{\textbf{6}} = \textbf{1} \oplus \textbf{35}$. The corresponding NN bond operators of the Hamiltonian read
\begin{eqnarray}
{\cal H}_{\bf 6\bar 6}^b &=& \mp (\frac{25}{4} \mathcal{P}_{\bf 1} +  \frac{1}{4} \mathcal{P}_{\bf 15\oplus 20'}) \nonumber\\
&=&  \mp (6 \mathcal{P}_{\bf 1} + \frac{1}{4} \mathbb{I})\,  ,  \label{eq:ham6bar6}\\
   {\cal H}_{\bf 66}^b &=& \mp \frac{\sqrt{2}}{2}(\frac{3}{4} \mathcal{P}_{\bf 15} -  \frac{5}{4} \mathcal{P}_{\bf 20'\oplus 1}) \nonumber \\
  &=&  \mp\frac{\sqrt{2}}{2} (2 \mathcal{P}_{\bf 15} - \frac{5}{4} \mathbb{I})    \label{eq:ham66} \\
  &=&  \pm\frac{\sqrt{2}}{2} (\mathbb{P}_{\mathrm{SU}(6)} + \frac{1}{4} \mathbb{I})  \,  , \label{eq:perm}
  \end{eqnarray}
where the $-$ and $+$ signs in (\ref{eq:ham6bar6}) and (\ref{eq:ham66}) correspond to the antiferromagnetic (AF) and ferromagnetic (F) couplings respectively. The two-site SU(4) Hilbert spaces spanned by $\bf 15$, $\bf 15\oplus 20'$ and $\bf 20'\oplus 1$ can be mapped on the spaces spanned by $\bf 15$, $\bf 35$ and $\bf 21$ of SU(6), respectively. Equations~(\ref{eq:ham6bar6}) and (\ref{eq:ham66}) are defined in terms of the alternating and uniform fundamental representation $\tiny\yng(1)$ of SU(6), respectively. In the following, we shall refer to these enhanced symmetry points  as 
SU(6) $\bf 6\bar6$ and SU(6) $\bf 6 6$ symmetric points. At the latter higher symmetry point, the bond operator can be written in terms of the SU(6) color permutation $\mathbb{P}_{\mathrm{SU}(6)}$, as shown in Eq.~(\ref{eq:perm}).

\begin{figure}
	\centering
\includegraphics[width=0.5\textwidth]{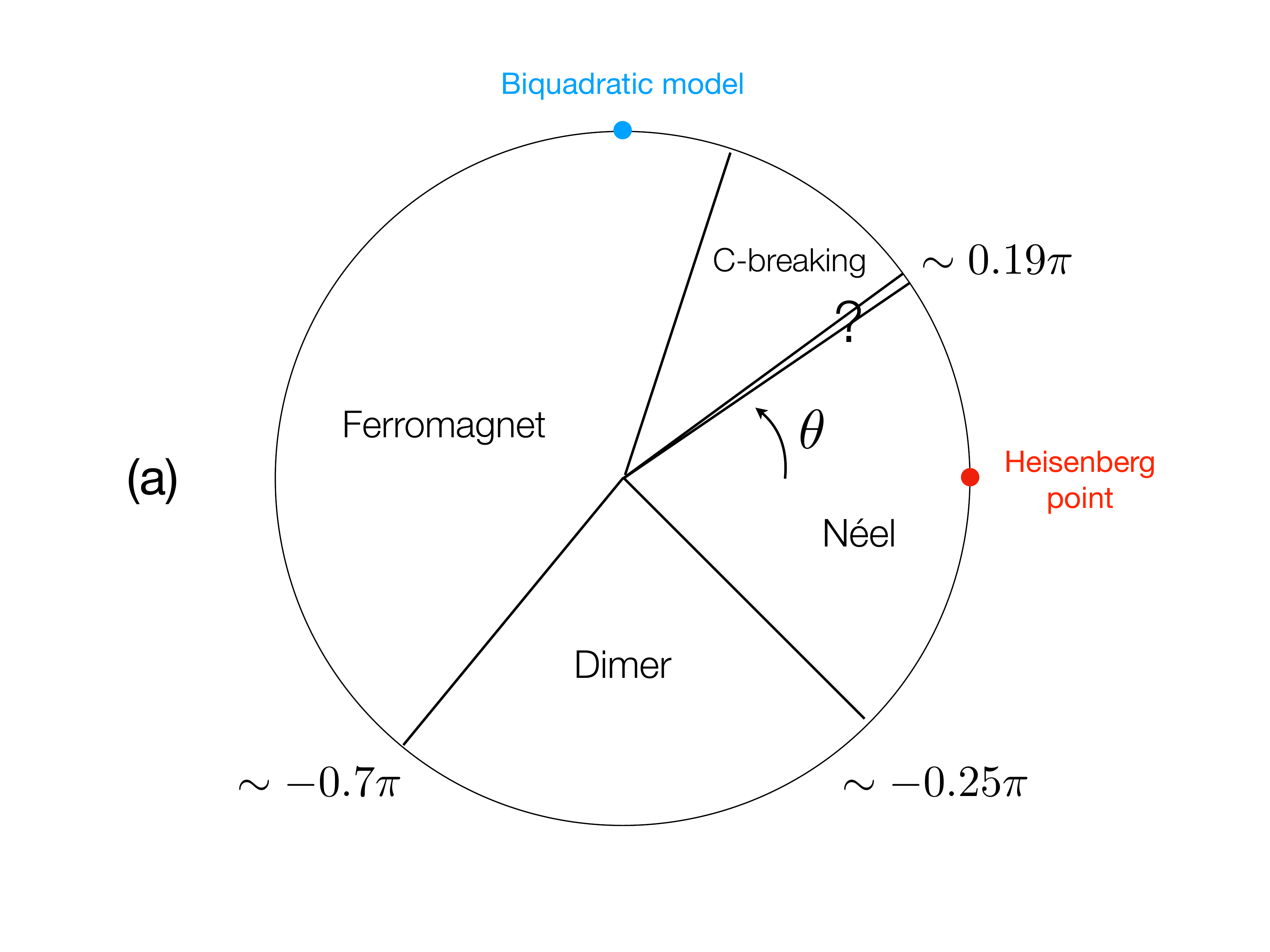}
\includegraphics[width=0.5\textwidth]{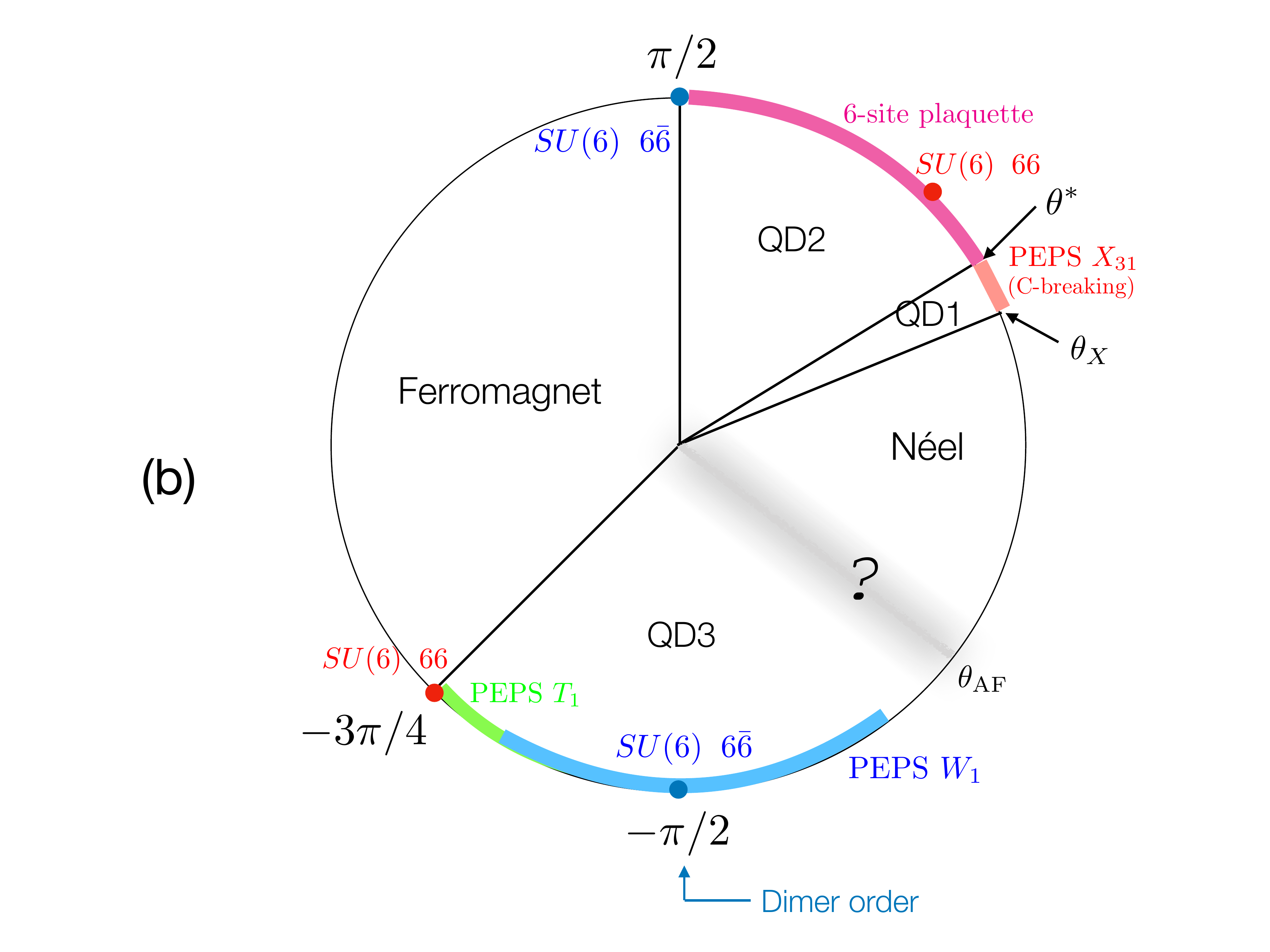}
	\caption{\footnotesize{Tentative phase diagrams as a function of $\theta$. (a) Adaptation from Ref.~\cite{paramekanti_sun_2007}. Antiferromagnet and ferromagnetic orders are expected in some region around the antiferromagnetic and the ferromagnetic ($\theta=\pi$) points. A dimer phase (or more generally a VBC phase) is expected around $\theta=-\pi/2$. A C-breaking phase as well as a staggered flux state (indicated by a question mark) have been also proposed. 
			(b) Phase diagram drawn from ED results on periodic clusters (see text). The ferromagnetic phase is limited by first-order transitions exactly at the SU(6)-symmetric points at $\theta=-3\pi/4$ and $\theta=\pi/2$ (showing massive level crossings on all finite clusters). 
We have identified 3 quantum disordered (QD) regions from low-energy singlet excitations and marked  
the regions where the PEPS QSL constructed in this work may be relevant.  The variational (SU(6)-symmetric) 6-site plaquette phase is also shown.
}}
	\label{fig:phase_diag}
\end{figure}

\section{Critical discussion of the phase diagram}

In this section we discuss the current understanding of the model. We start by drawing a tentative phase diagram based on the work by Paramekanti {\it et al.}~\cite{paramekanti_sun_2007}. We then discuss our ED  studies that bring new insights, still leaving a number of open issues. 

A pure bilinear model $\theta=0$ is expected to stabilize an ordered antiferromagnetic (N\'eel) phase that breaks SU(4) symmetry~\cite{kim_linear_2017}, similarly to SU(2) antiferromagnetic Heisenberg models. Similarly, a ferromagnetic phase is expected in the vicinity of the ferromagnetic Heisenberg point at $\theta=\pm\pi$. We build our work starting from early calculations based on projected wavefunctions  \cite{paramekanti_sun_2007}. A schematic phase diagram based on this approach is shown in Fig.~\ref{fig:phase_diag}(a). Interestingly, besides the expected magnetic phases mentioned above, their phase diagram shows SU(4)-invariant quantum disordered (QD) phases, a dimerized phase and a C-breaking phase.
It also suggests the existence of a third QD phase in a narrow region around $\theta = 0.19 \pi$ (thus for a sign of the biquadratic interaction appropriate to a half-filled fermionic SU(4) Hubbard model \cite{wang_competing_2014}), which they attribute to a gapless staggered flux state~\cite{affleck_su2n_1991}.  

\begin{figure}
	\centering
	\includegraphics[angle=0,width=0.5\textwidth]{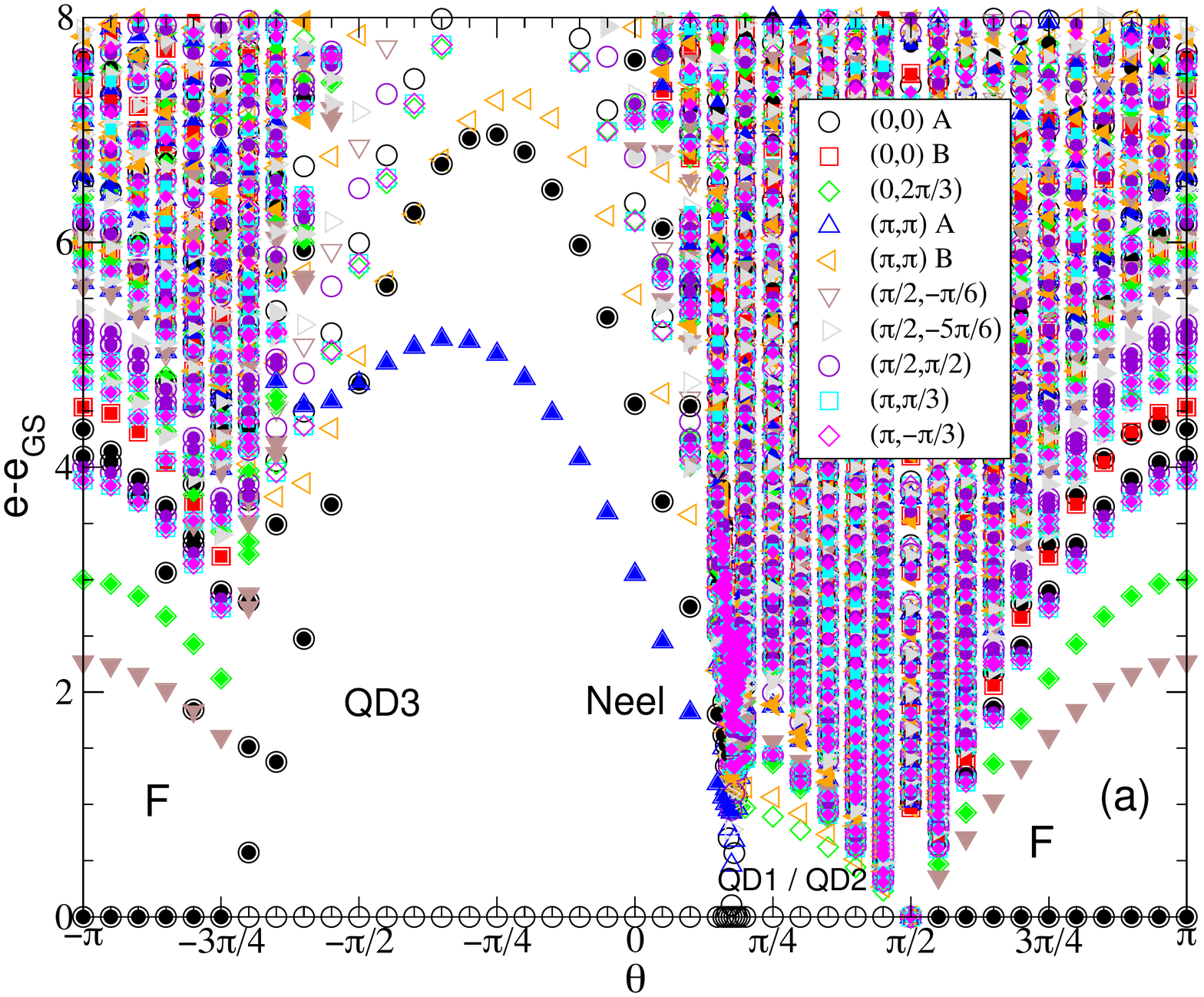}
	\includegraphics[angle=0,width=0.5\textwidth]{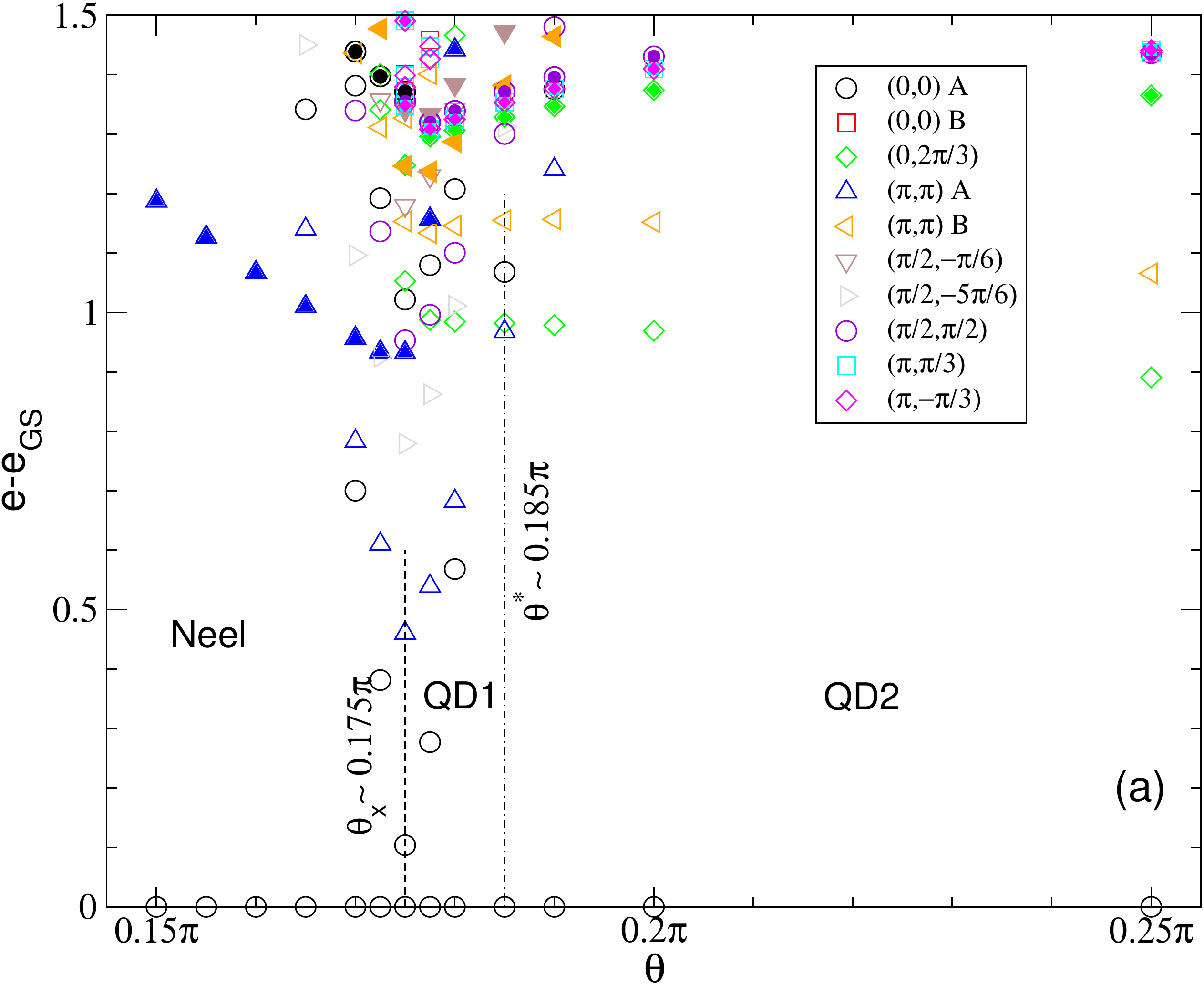}
	\caption{\footnotesize{ED energy spectrum of the periodic 12-site cluster. Open (closed) symbols correspond to singlets (higher dimensional irreps). Different symbols are used for different momenta in the reciprocal space. (a) Full parameter range $-\pi\le\theta\le\pi$. (b) Zoom of the range $0.15 \pi\le\theta\le0.25\pi$. Crossing of ground-state levels is observed at $\theta_X\sim 0.175\pi$, signaling a first order transition. Level crossings of (singlet) excited states at $\theta^*$ are marked by a dashed line.
	}}
	\label{fig:ed12}
\end{figure}

We have tried to refine the phase diagram using ED of four periodic $N=8,10,12,16$ square clusters, see Appendix~\ref{app:ed}. Note that these clusters unfortunately, have different lattice symmetries: for instance, reflection symmetry is missing in the 10- and 12-site cluster, the 16-site clusters can be mapped on  a 4-dimensional cube with larger symmetry,  and the reciprocal space of the 10- and 12-site clusters does not contain the ${\bf q}_{\rm VBC}=(\pi,0)$ wavevector.
However, all clusters show consistently the existence of two first order transitions characterized by the simultaneous crossing of 
many non-singlet SU(4) states (including the highest-weight multiplet of SU(4)) with a SU(4) singlet state, and occuring 
at exactly the SU(6)-symmetric points  $\theta=-3\pi/4$ and $\theta=\pi/2$, as shown in Figs.~\ref{fig:ed12}(a) and \ref{fig:eps_theta}(a). At these two crossings we have checked that the ground state correspond exactly to the SU(6) irrep of largest weight, i.e.\ to the SU(6) ferromagnet. These points hence mark the exact boundaries of the ferromagnetic phase as represented in the new phase diagram on Fig.~\ref{fig:phase_diag}(b).  The existence of a magnetic N\'eel phase is reflected by a ${\bf q}=(\pi,\pi)$ magnetic low-energy excitated state (i.e.\ belonging to a finite dimensional irrep) above the singlet GS. Due to finite-size effects, its precise boundary on one side is not fully accurate, as indicated by a question mark in Fig.~\ref{fig:phase_diag}(b). On the other side, we think it is limited by a very sharp level anti-crossing at $\theta_X\sim 0.175\pi$ as shown in Fig.~\ref{fig:ed12}(b) and more clearly in Fig.~\ref{fig:eps_theta}(c).
In all clusters, we see a narrow region around $\theta_X < \theta < \theta^*\sim 0.185\pi$ characterized by a few low-energy singlets with different momenta (see Fig.~\ref{fig:ed12}(b)) -- named QD1 in Fig.~\ref{fig:phase_diag}(b) -- that may be consistent with a QSL like e.g.\ the gapless staggered flux state or a C-breaking phase~\cite{affleck_su2n_1991}. The two phases at the boundary of the ferromagnetic region are more difficult to characterize. The 12-site cluster suggests the existence of two quantum disordered phases -- named QD2 and QD3 in Fig.~\ref{fig:phase_diag}(b) -- as signalled by a singlet GS with low-energy singlet excitation(s). In fact, $\theta=-\pi/2$ corresponds to the SU(6) ${\bf 6\bar 6}$ (antiferromagnetic) Heisenberg point whose GS is known to be dimerized, although with a quite small order parameter~\cite{harada_neand_2003}. At the SU(6) $66$  (antiferromagnetic) Heisenberg point at $\theta=\pi/4$ the ground state is always a non degenerate SU(6) singlet on clusters whose number of sites is multiple of 6.

The topological PEPS of Ref.~\cite{gauthe_su4_2019} is a priori a good candidate for QD phases, but other alternatives exist. In fact, it has been proposed that the QD2 phase spontaneously breaks C-symmetry in contrast to the PEPS ansatz of Ref.~\cite{gauthe_su4_2019}. This has motivated us to construct other PEPS family allowing or not for spontaneous C-breaking. The existence of SU(6)-symmetric points in the (1D) parameter space is also greatly constraining the PEPS family by allowing it to be fine tuned to these higher symmetries. Using iPEPS techniques we have investigated the relevance of our PEPS spin liquids in some separate regions of the phase diagram (see Fig.~\ref{fig:phase_diag}(b)). Note that  Lieb-Schultz-Mattis-Affleck theorem is likely to apply for the $\tiny\yng(1)$ irrep of SU(6) or the $\tiny\yng(1,1)$ irrep of SU(4), corresponding to 1/6 and 1/2 fermionic filling, respectively.  SU(6) and SU(4) spin liquids are therefore expected to be topological -- with 6-fold and (at least) 2-fold degenerate groundstates, respectively -- or critical.

\section{SU(4)-symmetric PEPS families} 

\subsection{Simple SU(4)-symmetric PEPS}
\begin{figure}
 	\centering
 	\includegraphics[width=0.5\textwidth]{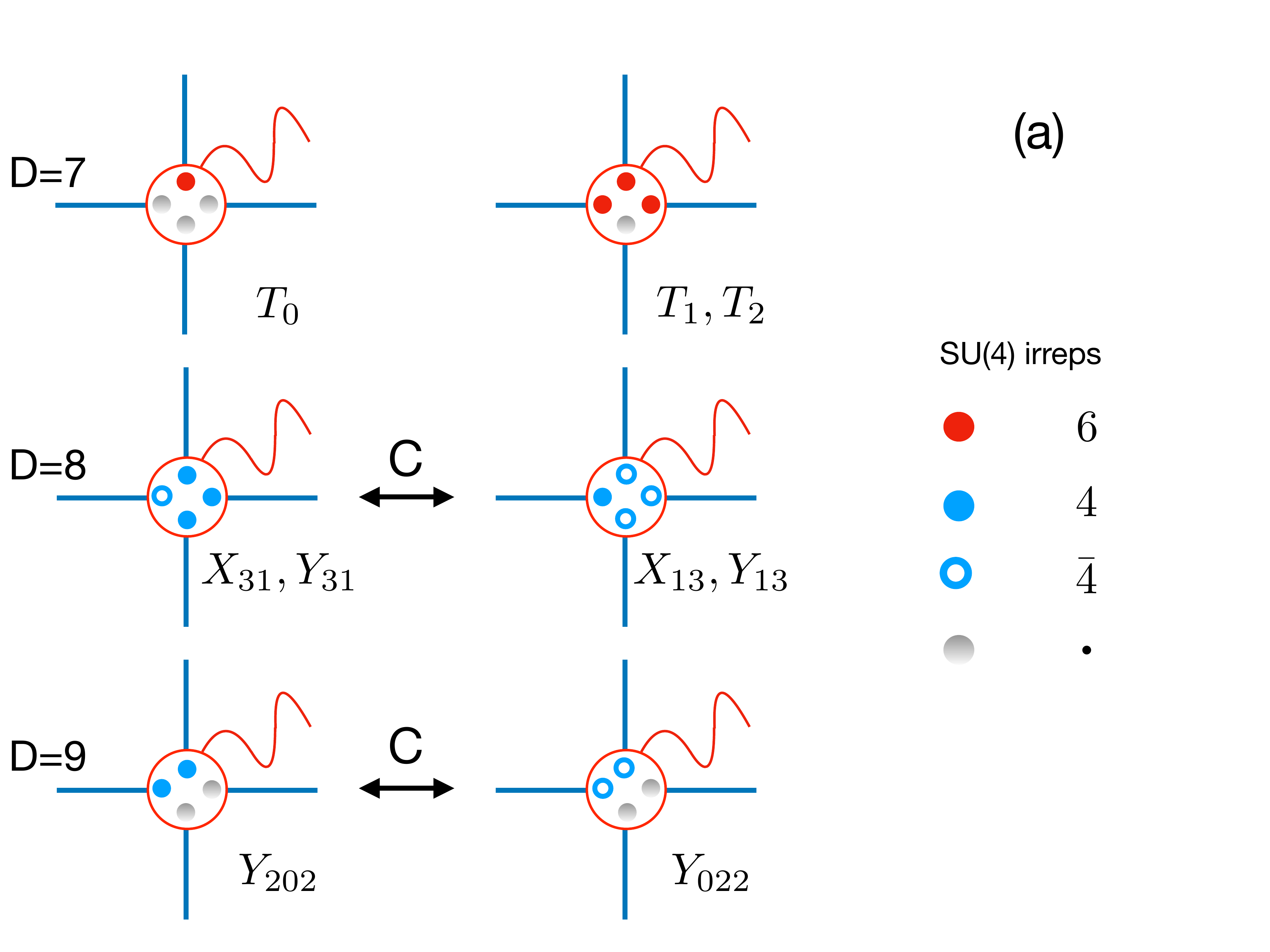}
 	\includegraphics[width=0.5\textwidth]{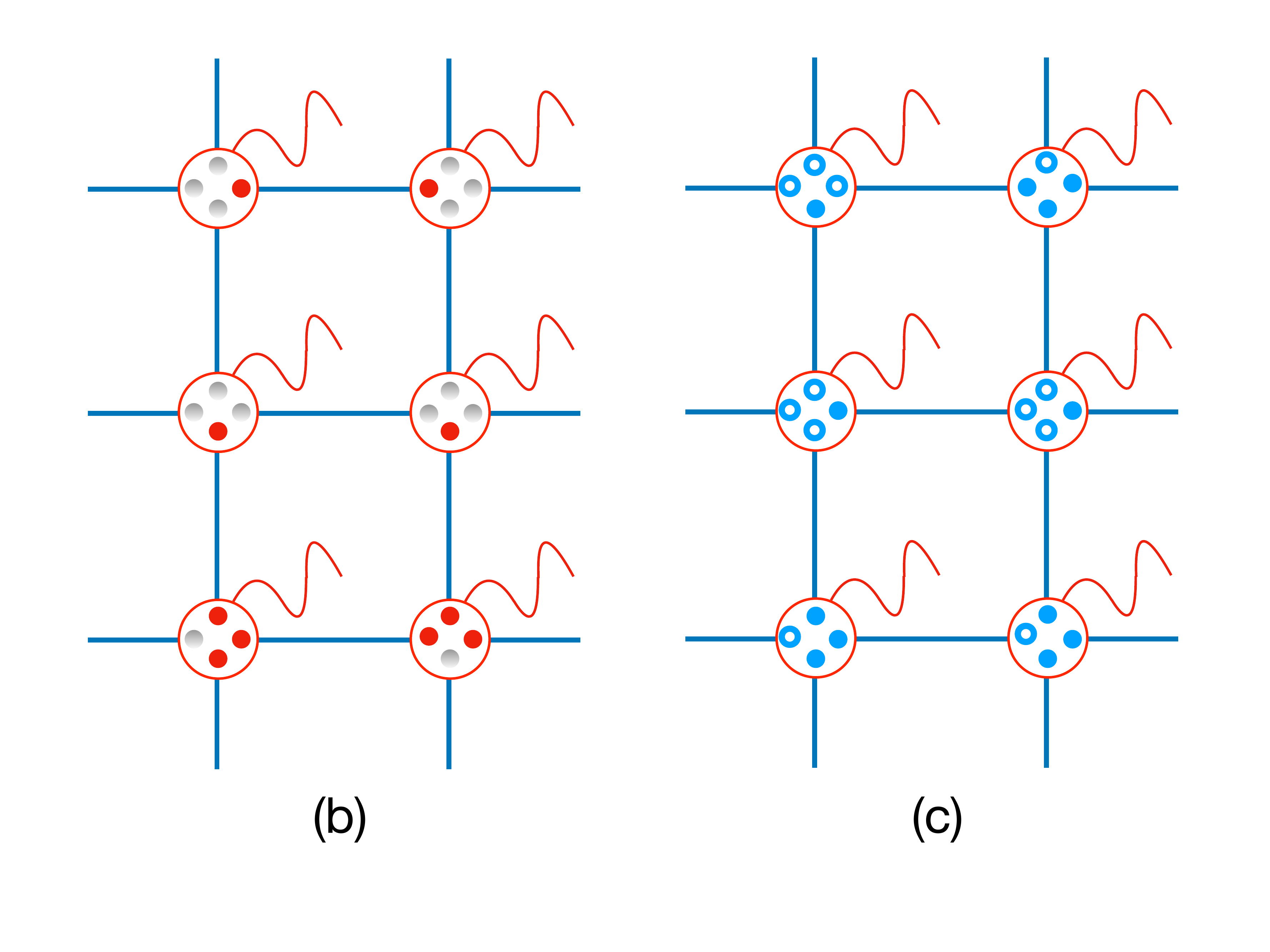}
 		\includegraphics[width=0.5\textwidth]{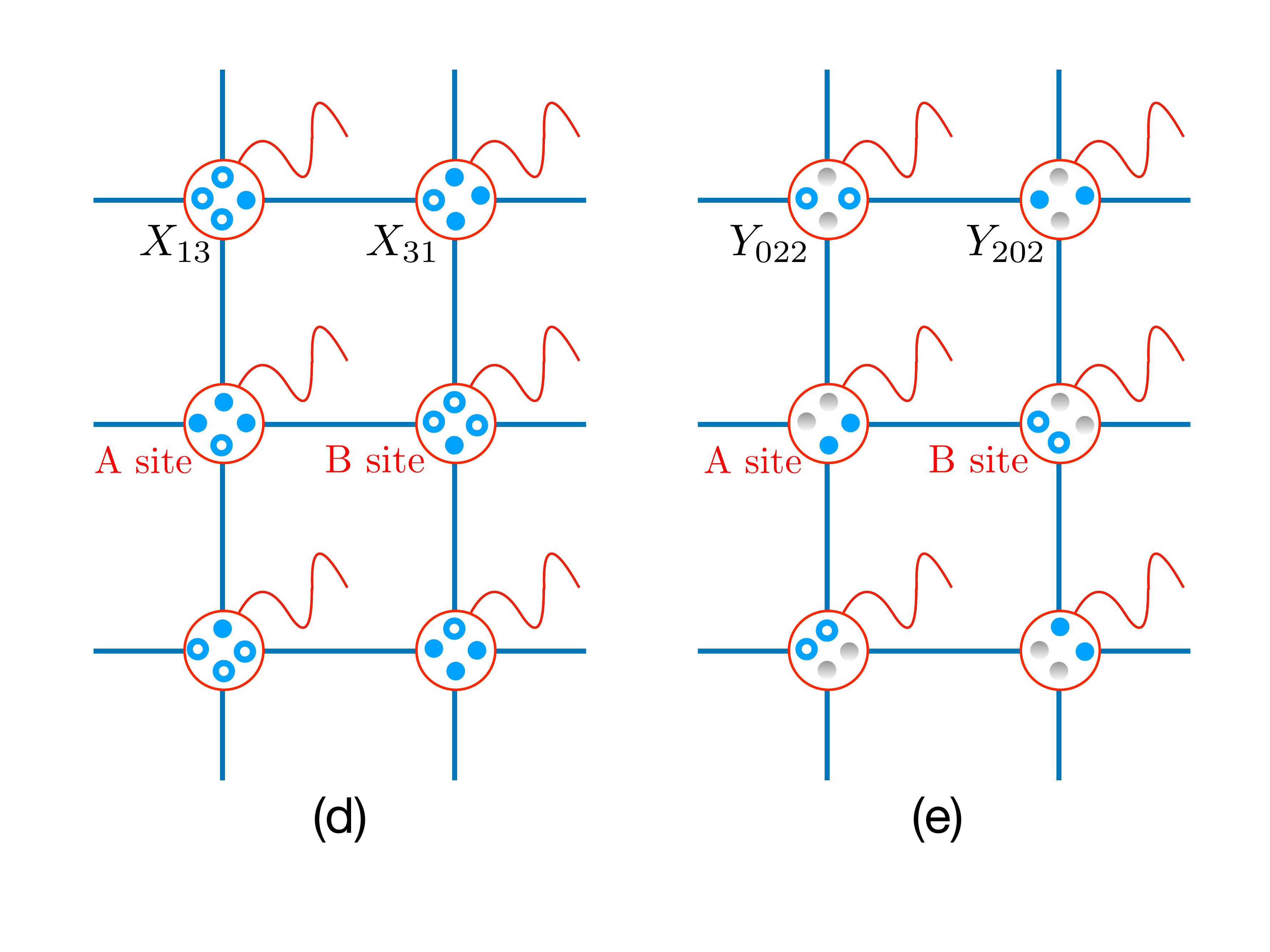}
 		\caption{\footnotesize{(a) The smallest $D$ SU(4)-symmetric tensors. The four virtual states (small dots) are projected (red circle) onto the physical $\bf 6$ irrep (wingly line). C-conjugated pairs of tensors are indicated by the arrows. (b) A typical PEPS configuration constructed from T tensors with either $\bullet\bullet$ or $\bf 66$ virtual singlet bonds. (c) A typical PEPS configuration constructed with $\bf 4\bar 4$ or $\bf{\bar 4}4$ virtual singlet bonds. (d,e) Typical PEPS configurations of C-breaking phases with a staggered arrangement of charge conjugated tensors. In panels (b), (c), (d) and (e), the matrices located on the bond centers that enforce the projections on SU(4) singlets are omitted for clarity. 
 	}}
 	\label{fig:TN}
 \end{figure}

We aim here to construct simple PEPS Ans\"atze on the square lattice which are fully invariant under SU(4) symmetry (ie the state is a global SU(4) singlet) and under all lattice symmetries (including lattice translations).
Our PEPS are defined by a single-site rank-5 tensor with four virtual indices on the $z=4$ bonds connecting the site to its neighbors and one index labeling the $d=6$ states of the physical $\tiny{\yng(1,1)}$ irrep 
as shown in Fig.~\ref{fig:TN}(a). The PEPS wavefunction is obtained by contracting the network of tensors on the virtual indices.~\footnote[1]{In fact, the pairs of nearest-neighbor virtual states are projected onto SU(4) singlets. This is enforced by inserting matrices on the bond centers. Examples of PEPS are shown in Fig.~\ref{fig:TN} (b), (c).}

To construct SU(4)-symmetric PEPS we follow here the framework developped in Ref.~\cite{mambrini_systematic_2016}. First, to 
enforce the invariance of the PEPS wavefunctions under 90-degrees rotation w.r.t.\ to any lattice site, the tensors should belong to 
the same one-dimensional  irrep of the point group $C_{4v}$, namely either to the rotation-even ${\cal A}_1$ or ${\cal A}_2$ irreps or to the rotation-odd  ${\cal B}_1$ or ${\cal B}_2$ irreps, where the subscripts 1 and 2 refer to even and odd characters w.r.t.\ axis reflections, respectively.
We shall not consider here the two-dimensional ${\cal E}$ irrep of $C_{4v}$. 
Hence, here after, we shall assume that the tensors belong to one of the four irreps of the point group, even if not explicitely specified. 
Secondly, to garanty (global) spin-rotation invariance, the virtual space $\cal V$ has to be a direct sum of SU(4) irreps (named ``species" or ``particles") in such a way that the 
expansion of ${\cal V}^{\otimes z}$ in terms of SU(4) irreps contains the physical irrep
 $\tiny{\yng(1,1)}$, possibly with some multiplicity. Restricting first to the smallest dimension $D=\text{dim}({\cal V})$, we are left with 
 \begin{eqnarray}
\cal{V}&=&\yng(1,1)\oplus\bullet\, \equiv \,{\bf 6}\oplus {\bf 1} \, , 
\label{Eq:61}  \\
\cal{V}&=&\yng(1)\oplus\yng(1,1,1)\,\equiv\, {\bf 4}\oplus\bar{{\bf 4}}\, , 
\label{Eq:44bar} 
\\
\cal{V}&=&\yng(1)\oplus\yng(1,1,1)\oplus\bullet\,\equiv\, {\bf 4}\oplus\bar{{\bf 4}}\oplus{\bf 1}\, ,
\label{Eq:44bar1}
%\cal{V}&=&\yng(1,1)\oplus\yng(1,1)\oplus\bullet\, \equiv \,{\bf 6}\oplus {\bf 6}^*\oplus {\bf 1} \, , 
%\label{Eq:661}  
 \end{eqnarray}
 with bond dimension $D=7$, $D=8$ and $D=9$, respectively. The different classes of tensors are shown in Figs.~\ref{fig:TN} (a).
 % and the exact expressions of all tensor elements can be found in  Appendix (\ref{app:tensors}). 
 Note that the $D=8$ tensors are just a subset of the set of $D=9$ tensors. Both $D=7$ and $D=8$ tensors have a $\mathbb{Z}_2$ {\it gauge} (ie connected to the virtual space only) symmetry since each of the two species entering $\cal V$ appears an odd number of times (1 or 3 times) on the 4 tensor {\it virtual} legs. 
 Importantly, we note in (\ref{Eq:44bar}) and (\ref{Eq:44bar1}) the emergence of a charge conjugation symmetry C exchanging $\bf 4 \leftrightarrow \bar 4$ and leaving the physical space $\bf 6$ invariant.  Note that unlike the SU(2) case, SU(4) charge conjugation is not a group operation.
 PEPS associated to the tensors with e.g.\ virtual space (\ref{Eq:61}) and (\ref{Eq:44bar}) can be constructed by contracting over the virtual indices, as shown in Figs.~\ref{fig:TN}(b) and \ref{fig:TN}(c), respectively. 
 
Let us first look more closely at the $D=7$ PEPS family: its generic on-site tensor $A_T$ is given by a linear combination of 3 (real) $D=7$ tensors $T_0$, $T_1$ and $T_2$ given in Ref.~\cite{gauthe_su4_2019},
\begin{equation}
A_T=a_0 T_0 + a_1 T_1 + a_2 T_2 \, ,
\label{eq:T}
\end{equation}
with $a_i\in\mathbb{R}$ and $a_0$ can be fixed to 1. 
These tensors are real and invariant under all symmetry operations of the lattice point group (i.e.\ the $C_{4v}$ group) -- namely they belong to the ${\cal A}_1$ representation of the group -- so that all PEPS of the family preserve parity (P) symmetry. The tensors can be labeled by an ``occupation number'' $n_{\rm occ}$ specifying, for each species in the virtual space, its total number on the four legs. E.g.\ for $T_0$, for which one has two different species $\bf 6$ and $\bf 1$, $n_{\rm occ}=\{1,3\}$. For $T_1$ and $T_2$, $n_{\rm occ}=\{3,1\}$. Properties of the $T$ tensors are summarized in Table~\ref{tab:tens_sym441}.
Note that the tensor $T_0$ alone generates the nearest-neighbor SU(4) Resonating Valence Bond (RVB) state~\cite{gauthe_su4_2019}. Note also that it is also possible to add a pure imaginary tensor $i a_3 T_3$ to (\ref{eq:T}), breaking 
time-reversal symmetry (T), while preserving all lattice symmetries. 

\begin{table}
	\vskip 0.5truecm
	\centering
	\begin{tabular}{CCCCCC}
		\hline
		\hline
		\mathrm{A} & D & {\cal V} & \quad n_\mathrm{occ}\quad  & \quad {\text C} \quad & \quad  C_{4v} \quad\\
		\hline
		T_0 & 7 & \textbf{6}\oplus\textbf{1} & \quad \{1,3\}  & \quad \sym \quad & \quad {\cal A}_1 \quad\\
		T_1 & 7 & \textbf{6}\oplus\textbf{1} & \quad \{3,1\} & \quad \sym \quad & \quad {\cal A}_1 \quad\\
		T_2 & 7 & \textbf{6}\oplus\textbf{1} & \quad \{3,1\} & \quad \sym \quad & \quad {\cal A}_1 \quad\\
		T_3 & 7 & \textbf{6}\oplus\textbf{1} & \quad\{3,1\}  & \quad \sym \quad & \quad {\cal A}_2 \quad\\
		\hline
		X_{31} & 8 & \textbf{4}\oplus\overline{\textbf{4}} & \quad\{3,1\}  & \quad \nsym \quad & \quad {\cal A}_1 \quad \\
		Y_{31}^{(i)} & 8 & \textbf{4}\oplus\overline{\textbf{4}} &\quad  \{3,1\} \ & \quad \nsym \quad & \quad {\cal A}_2 \quad \\
		X_{13} & 8 & \textbf{4}\oplus\overline{\textbf{4}} & \quad \{1,3\}  & \quad \nsym \quad & \quad {\cal A}_1 \quad \\
		Y_{13}^{(i)} & 8 & \textbf{4}\oplus\overline{\textbf{4}} & \quad \{1,3\} & \quad \nsym \quad & \quad {\cal A}_2 \quad \\
		\hline
		Y_{202} & 9 & \textbf{4}\oplus\overline{\textbf{4}}\oplus\textbf{1}\quad &\quad \{2,0,2\}  & \quad \nsym \quad& \quad {\cal A}_2 \quad \\
		Y_{022} & 9 & \textbf{4}\oplus\overline{\textbf{4}}\oplus\textbf{1} \quad & \quad \{0,2,2\}  &  \quad \nsym \quad & \quad {\cal A}_2 \quad\\
		\hline
		W_{(i)} & 13 & \textbf{6}\oplus\textbf{6}\oplus\textbf{1} \quad & \quad \{2,1,1\}  &  \quad \sym \quad & \quad {\cal A}_1 \quad\\
		\hline
		\hline
	\end{tabular}
	\caption{Classification of ${\bf 6\oplus 1}$ and ${\bf 4}\oplus\overline{\textbf{4}}$ SU($4$)-symmetric tensors in terms of virtual space, action of SU(4) charge conjugation (C) and $C_{4v}$ irreps. Tensors $T_0$, $T_1$ and $W_1$ are more symmetric and are invariant under any virtual leg permutation. The occupation numbers of each virtual species on the four virtual bonds are shown within brackets.}
	\label{tab:tens_sym441}
\end{table}

Tensors with virtual space ${\bf 4}\oplus\bar{{\bf 4}}$ can also be classified according to their point group symmetry. As shown in Table~\ref{tab:tens_sym441}, one (two) pair(s) of C-conjugated $X_{13}$ and $X_{31}$ ($Y_{13}^{(i)}$ and $Y_{31}^{(i)}$, $i=1,2$) tensors have ${\cal A}_1$ (${\cal A}_2$) symmetry. A general PEPS ansatz preserving $C_{4v}$ symmetry can be obtained from a local tensor combining all $X$ tensors in the following way~:
\begin{eqnarray}
A_X &=& A_R + i A_I \, , \\
A_R &=& X_{31} + \alpha X_{13} \, , \nonumber\\
A_I &=& \sum_i (\beta_{31}^{(i)}  Y_{31}^{(i)} + \beta_{13}^{(i)}  Y_{13}^{(i)} ) \nonumber \, ,
\label{eq:A_XY}
\end{eqnarray}
where $A_I\ne 0$ would spontaneously break time-reversal symmetry. 
Similarly, one can use the $Y_{202}$ and $Y_{022}$ tensors (of $ {\cal A}_2$ point group symmetry) which include an extra spin singlet in the virtual space,
$$
A_Y=Y_{202} + \gamma Y_{022} \, .
$$
In general $A_X$ ($A_Y$) breaks charge conjugation except when $\alpha=\pm 1$ and $\beta_{31}^{(i)} = \pm  \beta_{13}^{(i)}$ ($\gamma=\pm 1$). 

In the following we shall use charge-conjugated tensors, $A_X$ and $\overline{A_X}$ (resp. $A_Y$ and $\overline{A_Y}$) on the two A and B sublattices. 
Configurations of such states for $A_I=0$ and $\alpha=0$ (resp. $\gamma=0$) are shown in Fig.~\ref{fig:TN}(d) (resp. Fig.~\ref{fig:TN}(e)).
In that case, by acting with charge conjugation on the (physical) B sites, one can rewrite the PEPS in terms of a unique $A_X$ (or $A_Y$) tensor on all sites.  While the tensor network is translation invariant, it must be emphasized that it is the common PEPS representation of two different wavefunctions that are only invariant under a translation by \emph{two} sites (the translation by a lattice unit vector being equivalent to the conjugation of the whole lattice). These wavefunctions are orthogonal as their odd and even combinations are eigenvectors of the unitary charge conjugation operator with different eigenvalues. A local order parameter for translation and charge conjugation breaking can be constructed \cite{paramekanti_sun_2007}. Note that closely related (but non-equivalent) ansatze can also be constructed using the {\it same} tensors $A_X$ or $A_Y$ on {\it both} sublattices but replacing the bond singlet projectors by singlet projectors involving  the 8 virtual particles around each site of a given sublattice (so-called Projected Entangled Simplex States or PESS), following closely the original construction by Affleck et al.~\cite{affleck_su2n_1991} . 

Note that all these $T$, $X$ and $Y$ tensors, taken individually, have an extended U(1) gauge symmetry so it is expected that their associated PEPS would have critical correlations.  In fact, as can be seen in Figs.~\ref{fig:TN}(d,e), resonances between configurations is obtained by on-site permutations of virtual states along closed loops, and the PEPS inherits its critical nature from that of quantum loop models on bipartite lattices~\cite{Pollmann2011}.
Combining tensors with different $n_{\rm occ}$ lead to a lower ${\mathbb Z}_2$ gauge symmetry (see Appendix~\ref{subsec:gauge}).

Next, we turn to the extension of these tensors so that they can accommodate the emergent higher SU(6) symmetry present at isolated points of the phase diagram. This brings severe constraints on the tensor construction and on the form of the virtual space. In fact, we readily see that the tensors (\ref{Eq:44bar}) and (\ref{Eq:44bar1}) should be excluded since there is no 4-dimensional irrep in SU(6) to map into.  Below we shall restrict to the $\bf 6\bar 6$ SU(6) symmetry, while the more involved case of the $\bf 66$ symmetry is left for the appendix.
 
 \subsection{PEPS with higher $\bf 6\bar 6$ SU(6) symmetry}
 \label{subsec:tensors6bar6}
 
 At $\theta=-\pi/2$, we know from previous ED that the GS is a non-degenerate SU(6) singlet.
 It is therefore legitimate to try to enlarge the SU(4)-symmetric PEPS family in order to capture the higher SU(6) $\bf 6\bar 6$ symmetry of the model at this fined-tuned point. In fact, enforcing the enlarged symmetry leads to strong restriction on the site tensor. First, we note that the $T_0$ tensor alone has SU(6) symmetry and the associated SU(4) RVB PEPS is in fact a SU(6) $\bf 6\bar 6$ RVB state, i.e.\ in the alternating fundamental irrep of SU(6). Therefore, we expect this PEPS should already give a reasonable approximate description of the GS at, or in the vicinity of, the {\it antiferromagnetic} SU(6) $\bf 6\bar 6$ point at $\theta=-\pi/2$, as we shall discuss below.

To enlarge the PEPS family beyond $T_0$, we shall require that the SU(4) tensors originate from a mapping of tensors (i) which are SU(6) symmetric and (ii) whose virtual space should only contain self-conjugate irreps or pairs of conjugate irreps of SU(6) (to be able to form virtual SU(6) singlets on the bonds). The smallest possible SU(6) virtual space ${\cal V}_6$ could therefore be  
 \begin{equation}
 {\cal V}_6=\yng(1)\oplus\yng(1,1,1,1,1)\oplus\bullet\, \equiv \,{\bf 6}\oplus {\bf\bar 6}\oplus {\bf 1} \, , 
 \label{Eq:661a}  
 \end{equation}
 of dimension $D=13$. The corresponding PEPS tensors represent the 28 SU(6) fusion channels $\bf 6\otimes6\otimes{\bar 6}\otimes 1\rightarrow 6$, of the four virtual states onto the physical state. Mapping to SU(4) would require a virtual space with irreps of the same dimensions,
  \begin{equation}
 \cal{V}=\yng(1,1)\oplus\yng(1,1)\oplus\bullet\, \equiv \,{\bf 6}\oplus {\bf 6}^*\oplus {\bf 1} \, , 
 \label{Eq:661b}  
 \end{equation}
 where the $\bf 6$ (self-conjugate) irrep occurs with multiplicity 2, ie with two ``colors''. Then, SU(6) charge conjugation C in (\ref{Eq:661a}) naturally translates here into color exchange $\cal C$, which can be viewed as an element of the SU(2)-color gauge symmetry. Note that it is convenient to use the conjugate $\bf 6^*$ irrep of SU(4) (equivalent to $\bf 6$ up to a basis change) which naturally maps onto the $\bf\bar 6$ irrep of SU(6).  
  At the higher symmetry point, the tensors on the  A and B sites are related by charge/color conjugation C/$\cal C$.  Hence, they have occupation numbers w.r.t.\ the $\bf 6$, $\bf \bar 6/6^*$ and $\bf 1$ species, $n_{\rm occ}=\{1,0,3\}$ ($T_0$) or $n_{\rm occ}=\{2,1,1\}$ ($W$) on the A sites, and  $n_{\rm occ}=\{0,1,3\}$ ($\overline{T_0}/T_0^*$) or $n_{\rm occ}=\{1,2,1\}$ (${\overline W}/W^*$) on the B sites. Later on, we shall consider the three W tensors, $W_1$, $W2$ and $W_3$, of ${\cal A}_1$ symmetry. Similarly to the tensors $T_0$ and $T_1$, the tensor $W_1$ is invariant under any virtual leg permutation and has an extended $S_4$ symmetry.
  
\begin{figure}
 	\centering
 	\includegraphics[width=0.5\textwidth]{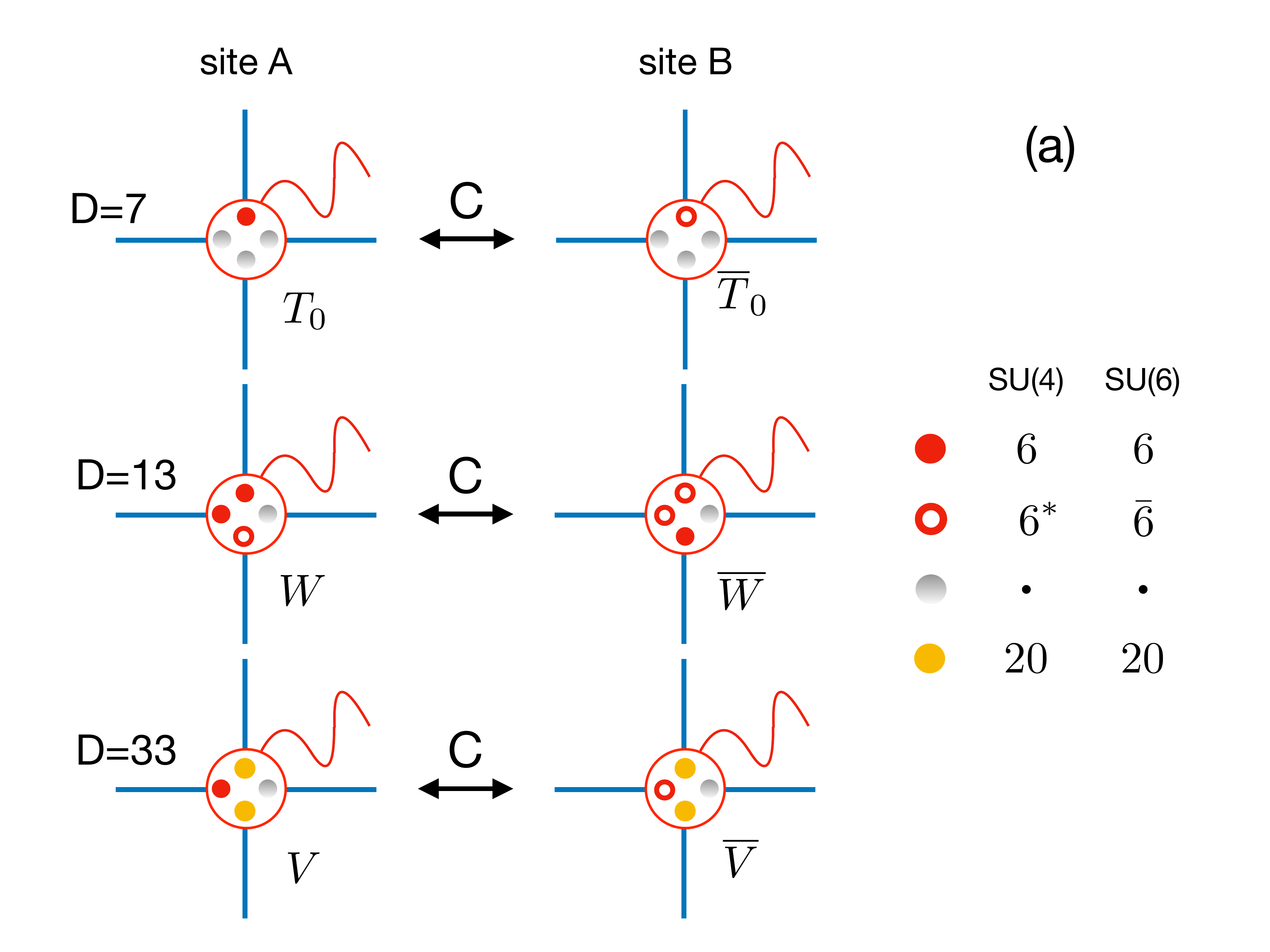}
 	%\vskip -0.1cm
 	%\includegraphics[width=0.5\textwidth]{tensors2b}
 	\vskip -0.2cm
 	\includegraphics[width=0.5\textwidth]{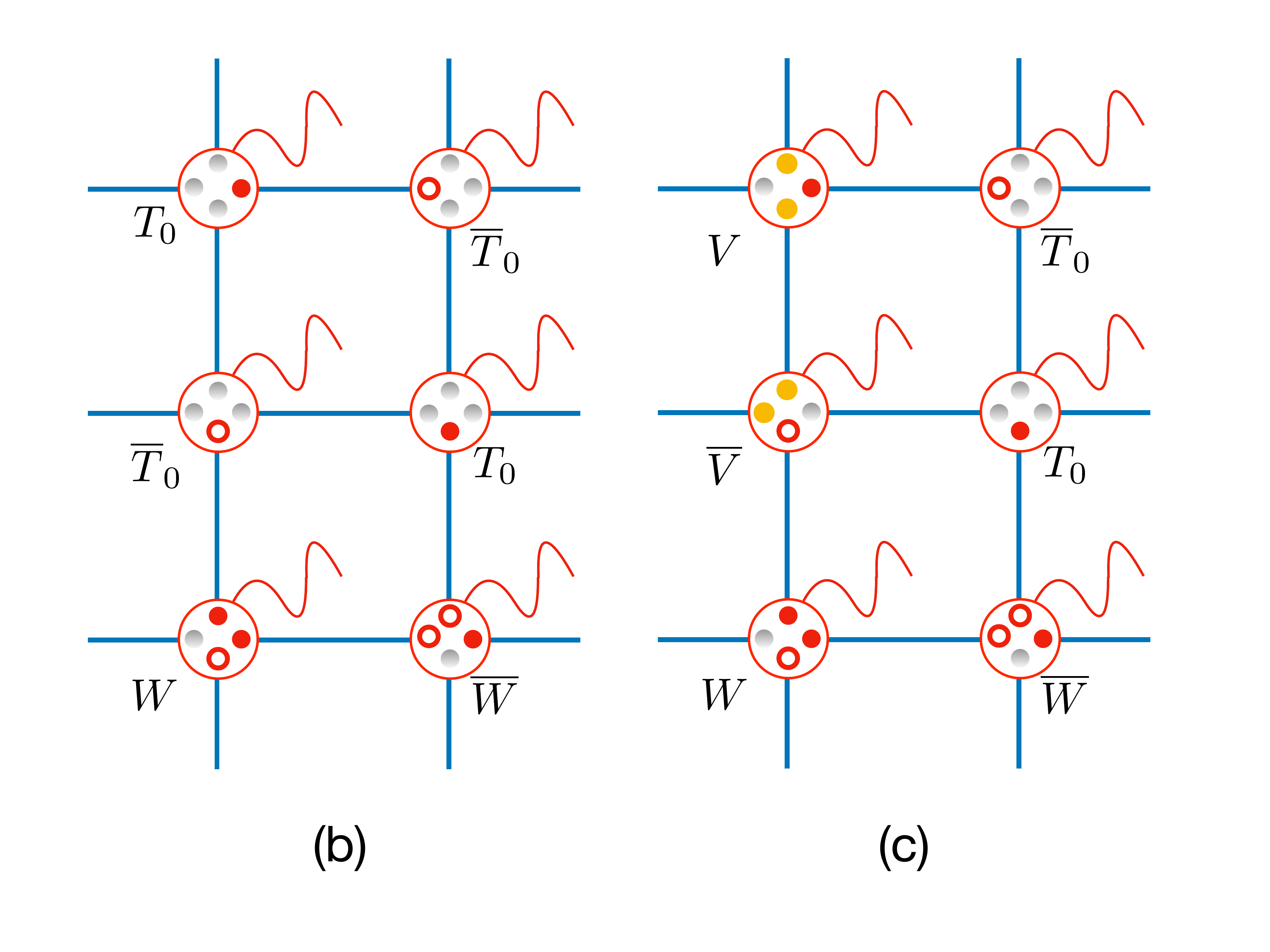}
 	\caption{\footnotesize{(a) Three classes of  SU(4) tensors with virtual spaces $\bf 6\oplus 1$,  $\bf 6\oplus 6\oplus 1 $ and  $\bf 6\oplus 6\oplus 20\oplus 1$ tensors, which can be mapped onto SU(6)-symmetric tensors  -- see Tables~\ref{tab:tens_occ1} and \ref{tab:tens_occ2}. Color-conjugated tensors have to be used on the A and B sites to encode the $\bf 6\bar 6$ SU(6) symmetry. A uniform SU(4) PEPS ansatz is obtained by performing a basis change on the (physical) B sites. (b) Configuration of a $\bf 6\bar 6$ SU(6)-symmetric  $D=13$ PEPS.   
 			A simpler PEPS involving only the $D=7$ tensors can be also constructed as shown on the 4 top sites.
 			(c) Configuration of a $\bf 6 \bar 6$ SU(6)-symmetric $D=33$ PEPS. 
 	}}
 	\label{fig:TN2a}
 \end{figure}

The SU(6)-symmetric PEPS is obtained by contracting the A and B tensors on all lattice bonds, as shown in Figs.~\ref{fig:TN2a} (b) and~\ref{fig:TN2a}(c). It is invariant under the combined action of C/$\cal C$ with a unit lattice translation. Hence, by acting with C/$\cal C$ on the (physical) B sites, one can rewrite the PEPS in terms of a unique tensor on all sites, making translation symmetry explicit.

One can even extend further the construction of the SU(6)-symmetric spin liquid by adding more irreps to the virtual space. The next step would be to include the self-conjugate (both in SU(4) and SU(6)) 20-dimensional irrep, giving $D=33$ virtual spaces,
\begin{equation}
{\cal V}_6=\yng(1)\oplus\yng(1,1,1,1,1)\oplus\yng(1,1,1)\oplus\bullet\, \equiv \,{\bf 6}\oplus {\bf\bar 6}\oplus {\bf 20}\oplus {\bf 1} \, , 
\label{Eq:66201a}  
\end{equation}
for SU(6) and,
\begin{equation}
\cal{V}=\yng(1,1)\oplus\yng(1,1)\oplus\yng(2,2)\oplus\bullet\, \equiv \,{\bf 6}\oplus {\bf 6}^*\oplus{\bf 20}^\prime\oplus{\bf 1} \, , 
\label{Eq:66201b}  
\end{equation}
for SU(4), as listed in Table~\ref{tab:tens_occ1}. Only three occupation numbers fits SU(6) fusion rules, defining one more $V$ class in addition to the two previous ones, as summarized in Table~\ref{tab:tens_occ2}. Since a large bond dimension $D=33$ is untractable with current algorithms, in the following we shall use, in addition to the $T$ tensors, the $W$ tensors with $D=13$.

\begin{table}
	\centering
	\begin{tabular}{CCC}
		\hline
		\hline
		%		\begin{align*}\begin{array}{ccc}
		\text{Occupation number} & \text{SU(4)} & \text{SU(6)} \\
		\hline 
		{\color{blue} \{0,1,0,3\} }& {\color{blue} 4 }& {\color{blue} 0} \\
		%	 \{0,1,0,3\} & 4 & 0 \\
		\{0,1,1,2\} & 12 & 0 \\
		\{0,1,2,1\} & 36 & 0 \\
		\{0,1,3,0\} & 40 & 0 \\
		{\color{brown} \{0,3,0,1\}} &{\color{brown}  12} & {\color{brown} 0} \\
		%	\{0,3,0,1\} & 12 & 0 \\
		\{0,3,1,0\} & 24 & 0 \\
		{\color{red}  \{1,0,0,3\}} & {\color{red} 4} & {\color{red} 4} \\
		%	\{1,0,0,3\} & 4 & 4 \\
		\{1,0,1,2\} & 12 & 0 \\
		\{1,0,2,1\} & 36 & 24 \\
		\{1,0,3,0\} & 40 & 0 \\
		{\color{blue} \{1,2,0,1\} }& {\color{blue} 36} & {\color{blue} 0} \\
		%	\{1,2,0,1\} & 36 & 0 \\
		\{1,2,1,0\} & 72 & 0 \\
		{\color{blue} \{2,1,0,1\} }& {\color{blue} 36 }& {\color{blue} 24} \\
		%	\{2,1,0,1\} & 36 & 24 \\
		\{2,1,1,0\} & 72 & 0 \\
		{\color{brown} \{3,0,0,1\} }& {\color{brown} 12} & {\color{brown} 0} \\
		%	\{3,0,0,1\} & 12 & 0 \\
		\{3,0,1,0\} & 24 & 0 \\
		{\color{green} 	\{0,2,1,0\} }&{ \color{green}0^* }& {\color{green} 12} \\
		\hline
		\hline
	\end{tabular}
	\caption{Full list of the classes of $D=33$ $\bf 6\oplus 6\oplus 20\oplus 1$ SU(4) symmetric tensors in terms of the occupation numbers of the four virtual particles. The second (third) column shows the number of SU(4) (SU(6)) fusion channels -- i.e.\ the total number of symmetric tensors (that could be further classified in terms of the irreps of the $C_{4v}$ point group, ${\cal A}_1$, ${\cal A}_2$, ${\cal B}_1$, ${\cal B}_2$ and ${\cal E}$). The $\bf 6\oplus 1$ $T_0$ ($T_1,T_2$) tensor(s) is (are) marked in red (brown).  The other $\bf 6\oplus 6\oplus 1$ tensors are marked in blue.  
		% Note that the total occupancy of the $\bf 6$ particles is odd, reflecting a gauge $\mathbb{Z}_2$ symmetry. 
		Note that the last tensor class (in green) belongs to the ${\cal V}=\bf 6\oplus 6\oplus 10\oplus \overline{10}\oplus 1$ family of SU(4) symmetric tensors. 
	}
	\label{tab:tens_occ1}
\end{table}

 \begin{table}
	\vskip 0.5truecm
	\centering
	\begin{tabular}{CCCC}
		\hline
		\hline
		\mathrm{Class} & D & {\cal V} & \quad n_\mathrm{occ\, [D=33]}\quad \\
		\hline
		T_0& \quad 7 \quad& \textbf{6}\oplus\textbf{1} & \quad \{1,0,0,3\} \quad \\
		\hline
		W& \quad 13\quad & \; \textbf{6}\oplus{\textbf{6}}\oplus\textbf{1} \; & \quad {\{2,1,0,1\}} \quad \\
		\hline
		V&\quad 33 \quad& \quad \textbf{6}\oplus{\textbf{6}}\oplus{\textbf{20}}\oplus\textbf{1} \quad &\quad  {  \{1,0,2,1\}} \quad   \\
		\hline
		\hline
	\end{tabular}
	\caption{SU(4)-symmetric tensors with $D\le 33$ accomodating higher $\bf 6\bar 6$ SU($6$) symmetry, classified according to their virtual space and occupation numbers in the $D=33$ largest SU(4) virtual space. 
	}
	\label{tab:tens_occ2}
\end{table}

\section{Tensor network algorithms}

The energy density (ie the energy per site) or local observables of our SU(4)-symmetric PEPS can be computed efficiently using a Corner Transfer Matrix Renormalization Group (CTMRG)  algorithm (see e.g.\ Ref~\cite{poilblanc_critical_2019} for details). It is based on a real space RG sheme, adding a single site at a time (starting from a corner) to construct an effective environment around some active region -- typically a small $2\times 1$ or $2\times 2$ cluster -- involving corner C and edge T ``fixed point" SU(4)-symmetric tensors. A parameter $\chi$ controls the amount of entanglement which is kept at each RG step in the Schmidt decomposition of the corner and, eventually, a $\chi\rightarrow\infty$ scaling is performed. Typically, $1/D^2$ of the largest Schmidt weights are kept at each stage. The CTMRG enables then to compute the energy densities $\epsilon[\theta,\{a_i\},\chi]$
of the various PEPS Ans\"atze in Hamiltonian (\ref{eq:H66}) (for each chosen $\theta$ value).

We then need to optimize the few coefficients $\{ a_i\} $ of the tensors to minimize the above PEPS variational energies. After starting from an initial arbitrary choice of the tensor coefficients, using CTMRG we obtain the converged environment tensors $C$ and $T$. We then evaluate the energy gradient numerically to ``feed" a Conjugate Gradient (CG) algorithm minimization routine which provides a new set of parameters. The procedure is repeated until the energy minimum is found. To take into account the error induced by the finite corner dimension $\chi$, we optimize the parameters for increasing $\chi$, starting from the previously optimized set of parameters. When a maximal value of $\chi$ is reached (imposed by computer power limitations), we use finite-entanglement scaling to extrapolate the energy in the $\chi \rightarrow \infty$ limit. 

The converged environment can then be used to compute the expectation value of any observable. It also allows to approximate the transfer matrix of an infinite one-dimensional (horizontal) strip obtained by contracting the TN in one (vertical) direction.
From the two largest eigenvalues $\lambda_1 > \lambda_2$ of this transfer matrix, one can compute the largest correlation length $\xi= 1/\ln(\lambda_1/\lambda_2)$ of the system. From its leading eigenvector, one can also obtain the environment entanglement entropy $S_\mathrm{Env}$, defined from a bi-partition of the infinite strip (see \cite{gauthe_su4_2019} for the technical details).

\section{Numerical results} 

\subsection{Ground-state and variational energies}

\begin{figure}
	\centering
	\includegraphics[angle=0,width=0.5\textwidth]{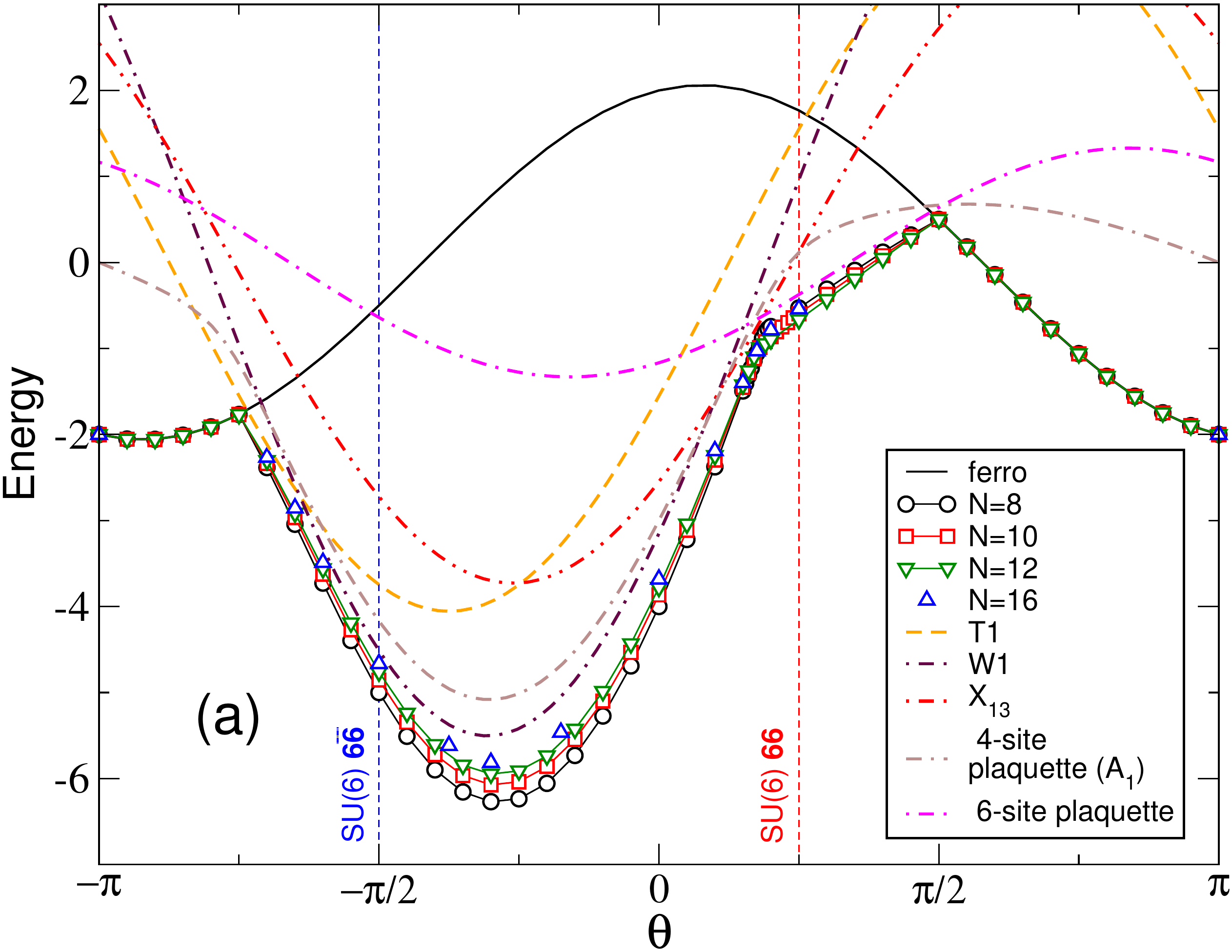}
	\includegraphics[angle=0,width=0.5\textwidth]{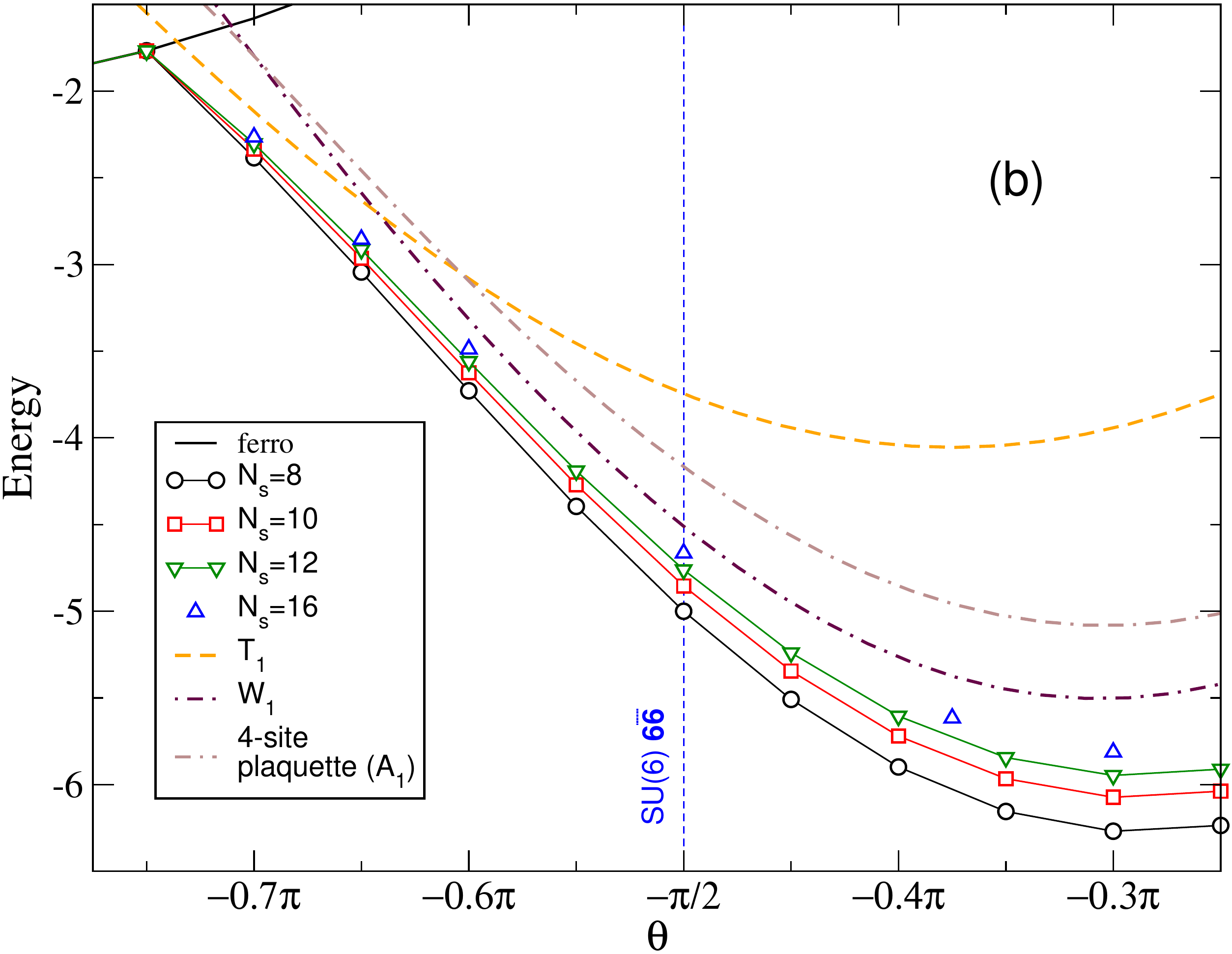}
	\includegraphics[angle=0,width=0.5\textwidth]{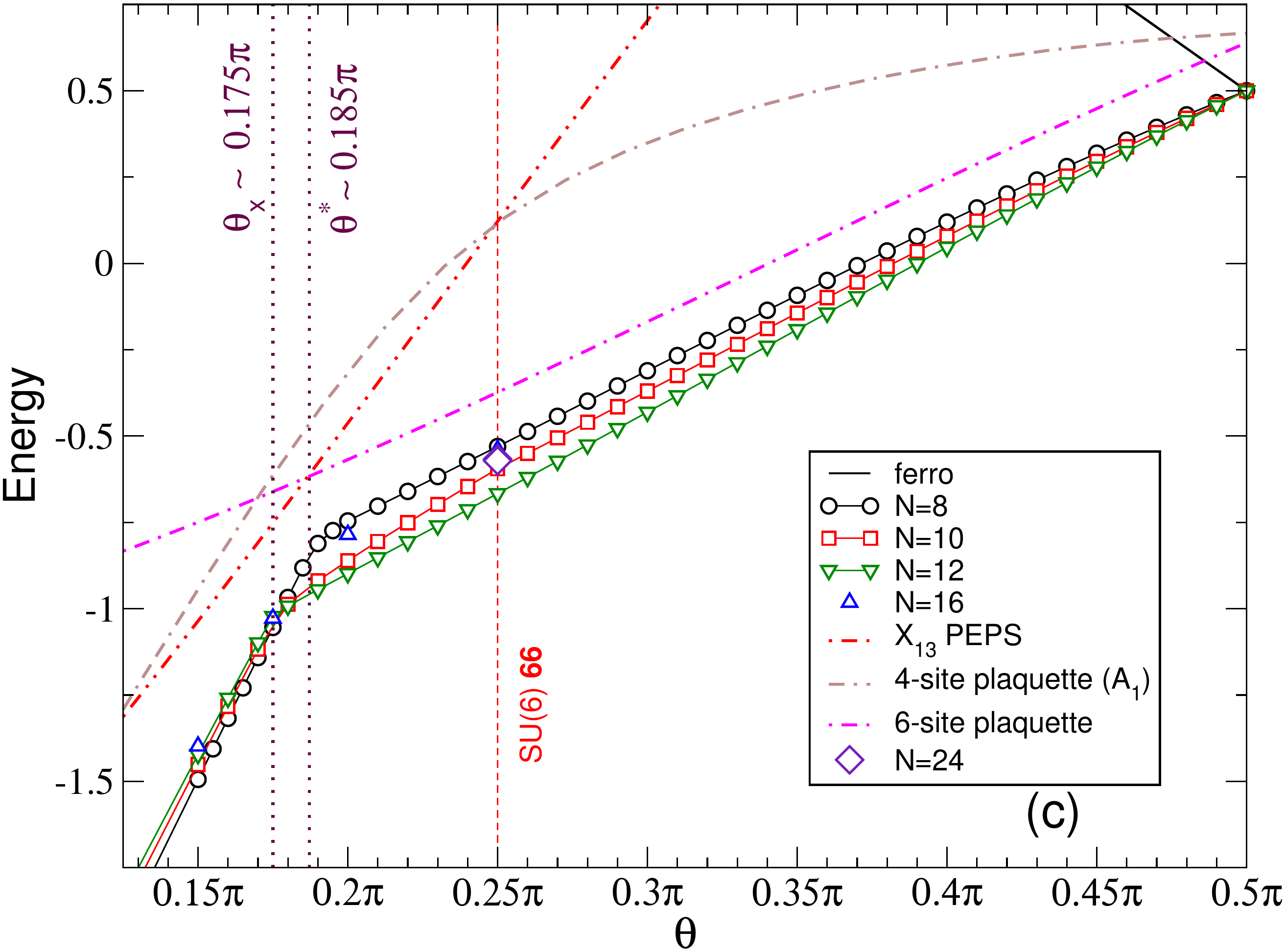}
	\caption{\footnotesize{Energies $\epsilon(\theta)$ of different wavefunctions under the Hamiltonian $\mathcal{H}(\theta)$. The PEPS energies are obtained after $\chi \rightarrow \infty$ extrapolation and compared to the  exact ferromagnetic state $\cal FM$, to 4-site and 6-site plaquette VBCs (see text) and to ED of small clusters.  The red points correspond to the optimization of the PEPS given by the $V=\textbf{6}\oplus \textbf{1}$ tensor (\ref{eq:T}). (a) Full parameter range $-\pi\le\theta\le\pi$. (b) Zoom of the range $-0.775 \pi\le\theta\le0.35\pi$. (c) Zoom of the range $0.125 \pi\le\theta\le0.5\pi$.
	}}
	\label{fig:eps_theta}
\end{figure}

In figure \ref{fig:eps_theta}(a), we have plotted, as a function of $\theta\in [-\pi,\pi]$, the energy density of some of the best low-energy PEPS, along with the finite size ED energies and the exact ferromagnetic and 4-site plaquette order wavefunctions (see Appendix \ref{sec:exact_wf} for details). As mentioned before, the ferromagnetic phase is exactly confined outside  of the range $-3\pi/4\le\theta\le\pi/2$. Although an exact expression for the energy of the quantum antiferromagnetic state is not known, the latter is believed to be stable in some extended region $\theta\in [\theta_{\rm AF},\theta_X]$ around $\theta=0$. The two regions on each side of this AF phase, $-3\pi/4 \le \theta \le \theta_{\rm AF}$ and  $\theta_X \le \theta \le\pi/2$ are likely to be SU(4)-symmetric quantum disordered phases of different nature, breaking or not lattice symmetry, and we examine them separately.

\subsection{Quantum disordered QD3 region}

\begin{figure}[h]
	\centering
		\includegraphics[angle=0,width=0.4\textwidth]{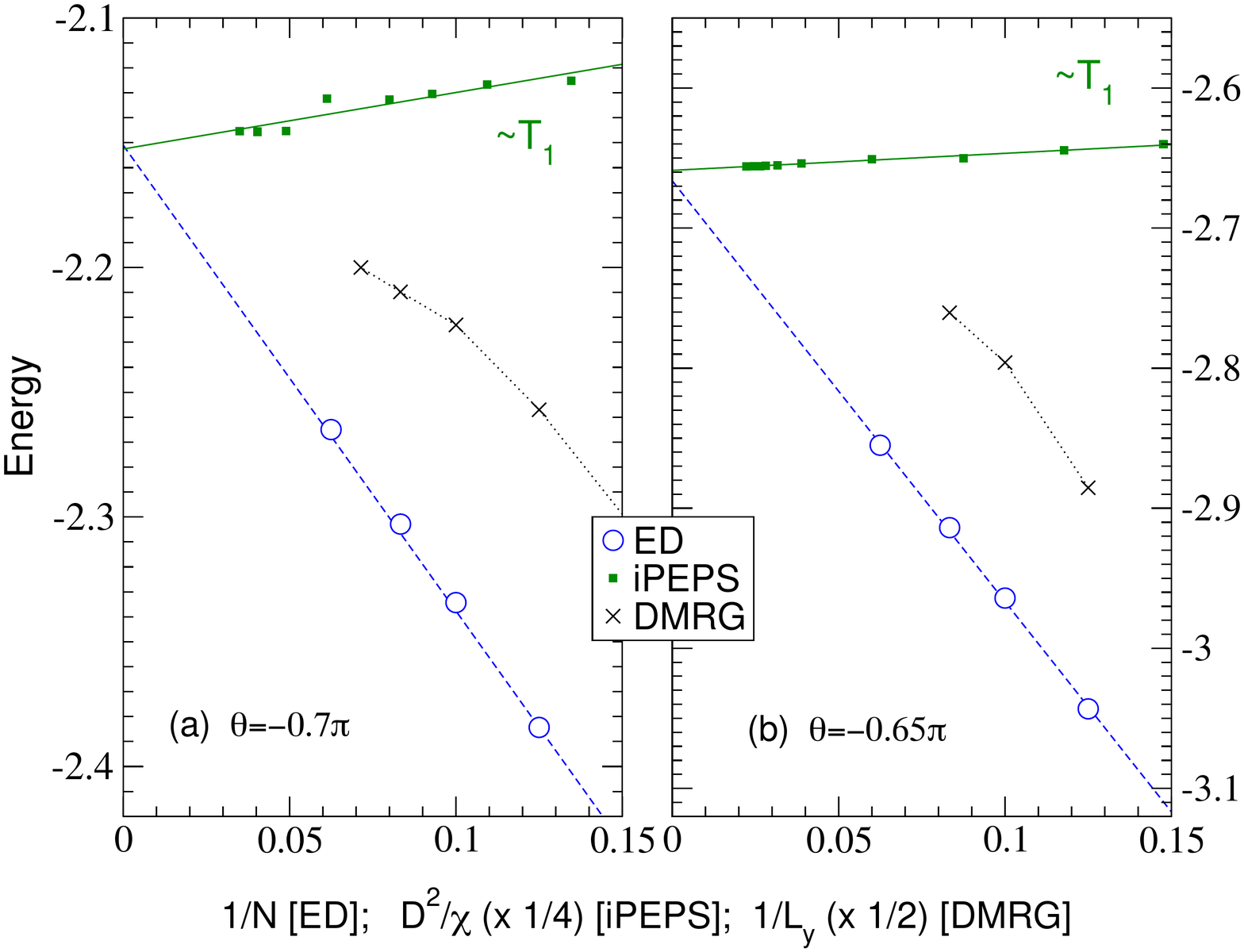}
		\includegraphics[angle=0,width=0.4\textwidth]{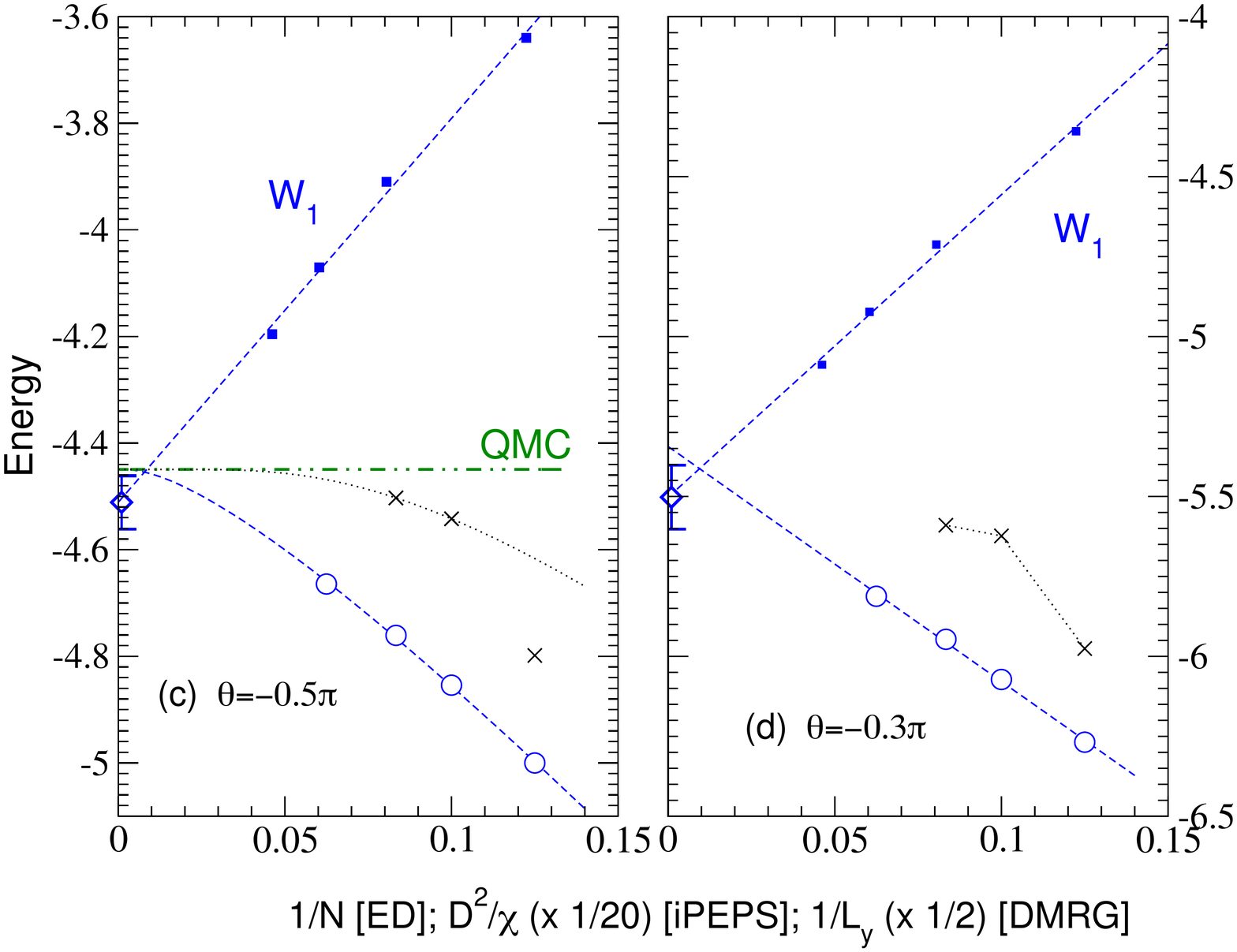}
		\includegraphics[angle=0,width=0.4\textwidth]{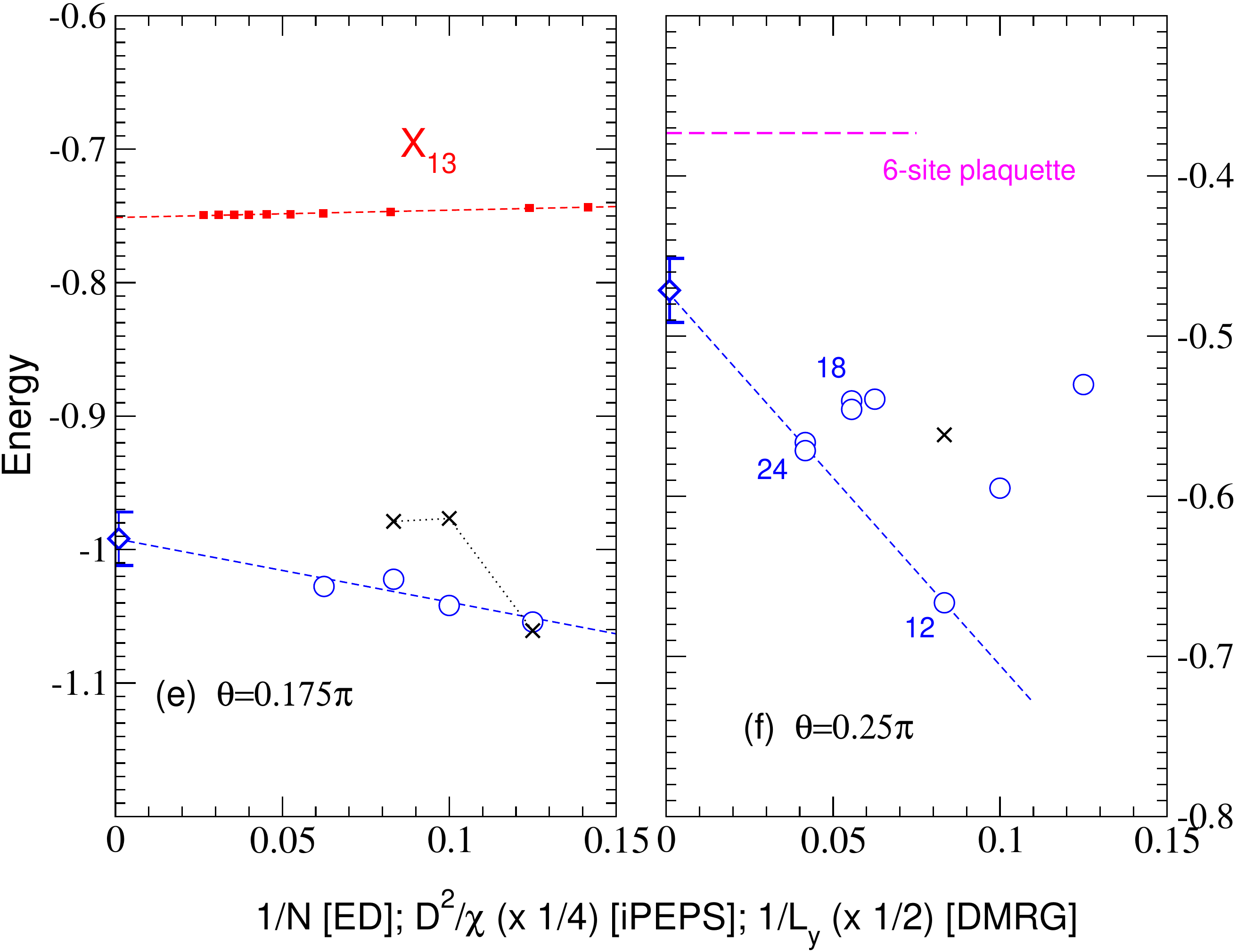}
		\caption{\footnotesize{Comparison of ED, DMRG and iPEPS energy densities for various values of $\theta$, as indicated on plots. Finite size (finite $\chi$) scaling is shown for ED periodic clusters and open DMRG cylinders (iPEPS infinite systems). Different PEPS are considered: $T_1$ (a,b), $W_1$ (c,d) and $X_{13}$ (e).  
		In (c), finite size scalings of the ED and DMRG data are attempted as $\epsilon=e_{\rm QMC}-ax\exp{-b/x}$, where $x=1/\sqrt{N}$ and $x=1/L_y$, respectively, and $e_{\rm QMC}$ is the (exact) QMC energy, providing estimates $\xi_{\bf 6\bar 6}=1/b$ of the correlation length $\sim 4.4$ and $\sim 5.4$. Otherwise, crude linear fits in $x$ are shown.
		The variational energy of the SU(6) 6-site plaquette phase is reported in (f). 
				}}
		\label{fig:eps_scaling}
\end{figure}

We first zoom in on the region $-3\pi/4 \le \theta \lesssim \theta_{\rm AF}$ in Fig.~\ref{fig:eps_theta}(b).
In this region, Paramekanti  and Marston  proposed a phase transition from the ferromagnetic phase to a dimerized phase. According to their VMC calculation the dimer order parameter of the best optimized (projected) dimerized state is close to fully saturated close to the transition with the ferromagnetic state. In other words, according to them, coupling dimers (within their variational manifold) does not decrease the energy. Although Paramekanti et al. do not quote any energy, we could nevertheless (approximately) estimate their best variational energy using a decoupled dimer mean-field solution (see Appendix \ref{sec:exact_wf}) and we found the energy of our uniform PEPS is significantly lower. This does not exclude that a small dimerization could take place but this definitely shows that our ansatz is better than Paramekanti's (projected) dimerized state when approaching the transition to the ferromagnet at $\theta=-3\pi/4$.
Here the optimum PEPS is obtained for $a_1\gg a_0,a_2,a_3$ (within the accuracy of our minimization for $\chi=D^2$) -- i.e.\ it is basically given by the $T_1$ tensor alone -- so it does not break time-reversal symmetry and may be critical (or have a very long dimer correlation length). As shown in figures \ref{fig:eps_scaling}(a) and \ref{fig:eps_scaling}(b) the finite-$\chi$ extrapolation of the optimum PEPS energy, performed for $\theta=-0.7\pi$ and $\theta=-0.65\pi$ respectively, are in excellent agreement with finite size scalings of ED and DMRG energies.~\cite{dmrg}
We therefore believe that this PEPS provides a good ground state candidate in this range of parameter and expect a QSL phase there.

Within this QD3 region, when moving towards the SU(6) $\bf 6\bar6$ point, the variational energy of the $A_T$ PEPS family deteriorates. However, as seen in figure \ref{fig:eps_theta}(b), we have found that the variational energy of the staggered PEPS constructed by putting the $W_1$ ($\overline{W_1}$)  $D=13$ tensor on the A (B) sites becomes remarkably good within a wide region around the SU(6) point. This is clear from the comparison of the finite-$\chi$ extrapolation of the PEPS energy with finite size ED and DMRG extrapolations, shown in figures \ref{fig:eps_scaling}(c) and \ref{fig:eps_scaling}(d) for $\theta=-0.5\pi$ and $\theta=-0.3\pi$ respectively. Early quantum Monte Carlo (QMC) accurate simulations at the SU(6) point (at which the minus-sign problem disappears) gave evidence for a dimerized phase, although with a quite small order parameter. In contrast, it is easy to see that our simple ansatz based on the (SU(6)-symmetric) $W_1$ tensor describes a non-degenerate translation invariant QSL. Nevertheless, due to a $\mathrm{U}(1)$ PEPS gauge symmetry, we expect its SU(6)-dimer correlations to be power-law so that this critical QSL can be viewed as a melted dimer-ordered state. Since the estimated correlation length of the dimer phase $\xi_{\bf 6\bar 6}\sim5.4$ is relatively long and the PEPS energy is remarkably close to the QMC energy (see figure \ref{fig:eps_scaling}(c)), we believe the latter gives a faithful representation of the ground state at not too long distances compared to $\xi_{\bf 6\bar 6}$. 
\subsection{Quantum disordered QD1 and QD2 regions} 

We now consider the region $\theta_X \sim 0.175\pi \leq \theta \leq \theta^*$ where $\theta^*$ is close to the value $\theta^*_{\rm 1D}=\arctan(2/3)\simeq 0.187\pi$, corresponding to an exact point in 1D. There, Paramekanti and Marston propose two possibilities, either a direct transition from a N\'eel phase to a charge-conjugation breaking phase or a thin QSL phase.
ED curves show a stark slope change around $\theta_X = 0.175\pi$, where we locate the transition from the AF state.

\begin{figure}
	\centering
	\includegraphics[angle=0,width=0.4\textwidth]{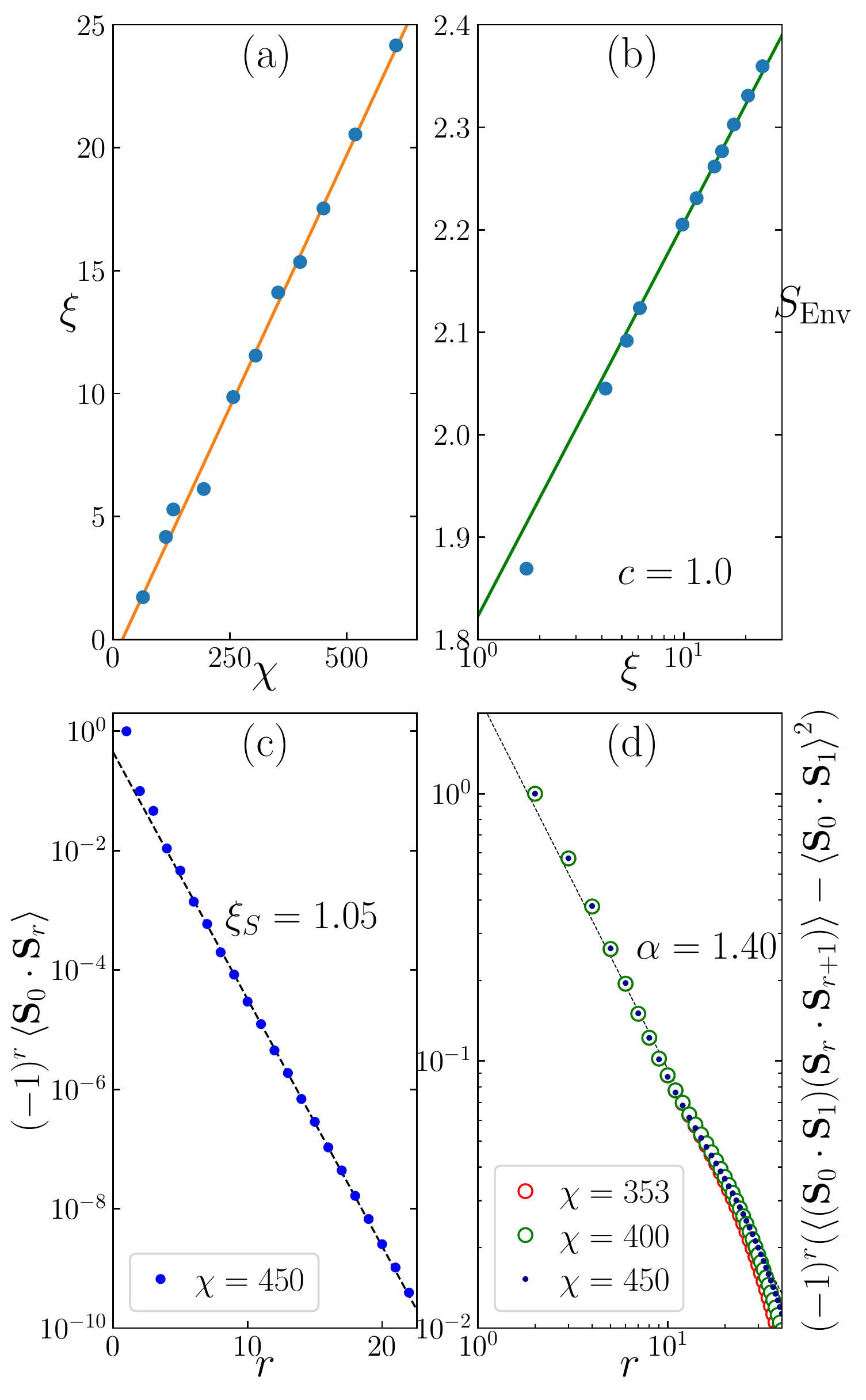}
	\caption{\footnotesize{Critical behavior of the $X_{13}$/$X_{31}$ PEPS ansatz. 
			(a) Largest correlation length $\xi$ vs $\chi$. (b) Scaling of the entanglement entropy vs the log of $\xi$. A central charge $c=1$ is proposed~\cite{calabrese_entanglement_2004}. (c) Two point (spin-spin) correlation function. An effective spin-spin correlation length is proposed with a fit $C_S(r+1) = C_S(1)\exp(-r/\xi_S)$ (d) Four point (dimer-dimer) correlation function, with critical exponent $\alpha$ fitted with $C_d(r+1) = C_d(2) r^{-\alpha}$.}
	}
	\label{fig:analysisX13}
\end{figure}

In our case, the tensors obtained from the $X$ and $Y$ family give a good energy in this region, a zoom on it is shown in figure \ref{fig:eps_theta}(c). Initial results obtained with the full form of equation (\ref{eq:A_XY}) show that charge conjugation is maximally broken in this region, therefore we thereafter restrict ourselves to $\alpha=\beta_{13}^{(i)}=0$. In the 1D case, a MPS obtained from a similar construction gives the exact ground state for $\theta=\theta^*_{\rm 1D}=\arctan(2/3)\simeq 0.187\pi$. 
The estimation of the energy of the $X_{31}$ PEPS (represented in figure \ref{fig:TN}(d)) is in reasonably good agreement with ED data as shown in figure  \ref{fig:eps_scaling}(e). 
Note that this PEPS build from a single site tensor bears some U(1) gauge symmetry so that we expect critical correlations. Indeed, figure \ref{fig:analysisX13}(a) shows that the largest correlation length (obtained from the transfer operator) increases linearly with $\chi$, with no sign of saturation. The scaling of the entanglement entropy $S_\mathrm{Env}$ w.r.t.\ the 
latter correlation length $\xi$ according to the formula $S_\mathrm{Env} = S_0 + c/6\log\xi$~\cite{calabrese_entanglement_2004} in \ref{fig:analysisX13}(a) suggests Conformal Field Theory (CFT) criticality with central charge $c=1$. 
Spin-spin and dimer-dimer correlations show very different qualitative behaviors, with exponential and algebraic decays, as shown in  figures \ref{fig:analysisX13}(c) and  \ref{fig:analysisX13}(d), respectively. While we expect the other PEPS (represented in figure \ref{fig:TN}(e)) obtained from the tensors $Y_{202}$ and $Y_{022}$ to be also relevant in this region (it is the simplest adaptation of the C-breaking phase construction of reference \cite{affleck_su2n_1991}), we were not able to compute an environment for these tensors. In any case, we expect a C-breaking phase that also breaks translation symmetry but preserves SU(4), as mimicked by our simplistic $X_{31}$ PEPS. Slightly lower energies reached with non-zero $\beta_{31}^{(i)}$ indicate this phase may also break time-reversal symmetry.

For $\theta^* \leq \theta \leq \pi/2$, our best results were obtained with uncorrelated, singlet plaquettes of 6-sites (see Appendix~\ref{sec:exact_wf}), which have better energies than our different PEPS Ans\"atze as shown in \ref{fig:eps_theta}(c). Such a plaquette can also be realized with the fundamental irrep of SU(6) on every sites and indeed no special behavior is observed at the SU(6) point $\theta=\pi/4$.
There is however a clear crossing of the energy curves at $\theta=\pi/2$, which corresponds to the transition to the FM state as discussed before. Hence, in this region, no evidence for a QSL phase is found, as the energies of our symmetry-preserving PEPS Ans\"atze are well above.

%%%%%%%%%%%%%%%%%%%%%%%%%%%%%%%%%%%%%%%%%%%%%%%%%%%%%%%%%%%%%%%%%%%%%%%%%%%%%%%%%%%%%%%%%%%%%%%%%%%%%%%%%
\section{Conclusion}

In this study, we have investigated the most general two-fermion SU(4)-symmetric Hamiltonian on the square lattice, with interaction limited to nearest-neighbor distance.
We combined ED, DMRG and PEPS techniques to propose a new phase diagram. We first argue the ferromagnetic domain is limited by the two SU(6) points at $\theta=\pi/2$ and  $\theta=-3\pi/4$. 
We then explore the rest of the phase diagram using different PEPS Ans\"atze and comparing them with DMRG and ED. 
We construct several SU(4) and $C_{4v}$ symmetric PEPS, depending on very few parameters, and explain how to construct tensor families  that can capture the extended symmetry at the SU(6) points. 

Previous QMC results on the SU(6) point $\theta=-\pi/2$ indicate a (weakly) dimerized phase there, and the purely bilinear point $\theta=0$ is known to belong to the AF phase.
Although the restriction to uniform symmetric PEPS prevents us from accessing dimerized and N\'eel phases, our PEPS still provide excellent variational energies in both i) an extended region from the ferromagnetic boundary at $\theta=-3\pi/4$ all the way to $\theta\sim -0.30\pi$ (including the SU(6) dimerized point) and ii) a narrower region around $\theta=0.18\pi$.
In the first region, due to the simplistic nature of the two low-energy best PEPS -- they are build from a (fixed) single tensor encoding a continuous U(1) gauge symmetry -- critical dimer correlations are found, which may be quite unstable~\cite{corboz_finite_2018,isakov_dynamics_2011}. Hence, minimal refinement of these PEPS wave functions could lead to either i) a short-ranged QSL with topological order -- by breaking the continuous gauge symmetry to a discrete gauge symmetry -- or ii) a weakly dimerized phase -- by allowing a two-sublattice modulation of the site tensor. Although, the second scenario agrees with the known physical behavior at the SU(6) point $\theta=-\pi/2$, the first scenario of a topological QSL may well be realized closer to the ferromagnetic phase transition at $\theta=-3\pi/4$. 
 
In the narrower region around $\theta\sim 0.18\pi$, we found no evidence for a {\it uniform} QSL but, rather, a $q=(\pi,\pi)$-modulated charge conjugation-breaking phase. Such a phase is a natural generalization of the 1D C-breaking phase, and has a particularly simple PEPS representation. Our analysis also suggests a transition from this phase to a 6-site plaquette phase, which seems to extend all the way to the ferromagnetic phase transition point at $\theta=\pi/2$. Of course, one cannot exclude that $\theta^*-\theta_x\rightarrow 0$, for increasing cluster sizes in ED, in which case one would observe a direct transition from the N\'eel state to the plaquette phase and no intermediate C-breaking phase. 

Lastly, we note that SU(4) qualitatively differs from SU(2) where no QSL arises in the case of NN interactions only. This leaves open the possibility of experimental realizations using any ultracold alkaline-earth atoms realizing SU($N$) symmetry by simply tuning the number of species~\cite{pagano_one-dimensional_2014} to e.g.\ $N=4$. Fixing a filling of two particles per site should avoid three-body losses and thus allow controlled experiments.

%%%%%%%%%%%%%%%%%%%%%%%%%%%%%%%%%%%%%%%%%%%%%%%%%%%%%%%%%%%%%%%%%%%%%%%%%%%%%%%%%%%%%%%%%%%%%%%%%%%%%%%%%
{\it Acknowledgments.}
This research was supported in part by the French Research Council (Agence Nationale de la Recherche, France) under Grants No. TNSTRONG ANR-16-CE30-0025 and No. TNTOP ANR-18-CE30-0026-01 and by the National Science Foundation under Grant No. NSF PHY-1748958.  This
work was granted access to the HPC resources of CALMIP
supercomputing   center   under   the   allocation   2018-P1231.
We  acknowledge  inspiring  conversations with Fabien Alet, Matthew Hastings, Fr\'ed\'eric Mila, Masaki Oshikawa, Karlo Penc,  and Keisuke Totsuka. We are also indebted to Pierre Nataf for providing the ED ground state energy (using the method of \cite{nataf_exact_2014}) of the SU(6) permutation model on 18 and 24 site clusters, as well as for a careful reading of the manuscript.
%

%%%%%%%%%%%%%%%%%%%%%%%%%%%%%%%%%%%%%%%%%%%%%%%%%%%%%%

\appendix
\section{Simple Ans\"atze}
\label{sec:exact_wf}

We consider here a few simple exact SU(4) states on the square lattice to compare to our PEPS. These states correspond to magnetic or quantum disordered phases, some of them introduced in Ref.~\cite{paramekanti_sun_2007}.
On a given lattice with coordination number $z$ and nearest neighbor coupling, the energy per site is one half of the average value of all the $\mathcal{H}(\theta)$ taken on the $z$ bonds. The energy of a given wavefunction for all $\theta$ is a sinusoid  parameterized by its value in two points only: $\expval{\mathcal{H(\theta)}} = \cos(\theta) \, \mathcal{H}(0) + \sin(\theta) \, \mathcal{H}(\pi/2)$.

\paragraph{Ferromagnetic state $\mathcal{FM}$.} This state is the most symmetric and any pair of sites in the lattice is projected in the most symmetric irrep $\textbf{20}$. In this state, $\langle \mathcal{H}(\theta)\rangle_\textbf{20} = \cos(\theta) + \frac{\sin \theta}{4}$ and, therefore, the energy per site is
\begin{equation}
e_\mathcal{FM} =  2\cos \theta + \frac{1}{2}\sin \theta
\end{equation}

\paragraph{Uncorrelated state $\mathcal{U}$.} In this state, each site is totally uncorrelated from its neighbors, which means $\langle \mathcal{H}(\theta)\rangle_\text{unc} = \text{Tr}(\mathcal{H})/36 = \frac{5}{12}\sin \theta$. Hence the energy per site is
\begin{equation}
\label{eq:uncorr}
e_\mathcal{U} = \frac{5}{6}\sin \theta
\end{equation}

\paragraph{Dimerized state $\mathcal{D}$.} In a fully dimerized state, each site belongs to one singlet of energy $\langle \mathcal{H}(\theta)\rangle_\textbf{1} = -5 \cos(\theta) + \frac{25}{4} \sin \theta$.  All the other neighbors are totally uncorrelated. Every dimer covering states have the same  energy per site
\begin{equation}
e_\mathcal{D} = -\frac{5}{2} \cos \theta + \frac{15}{4} \sin \theta
\end{equation}

\paragraph{4-site plaquette states $\mathcal{P}_4$.} 
On the square lattice we can also construct states where four sites in a square form a singlet and cover the lattice with these plaquettes. This state spontaneously breaks the translation invariance of the lattice. To construct it, we have to consider the projector $\tiny{\yng(1,1)}^{\otimes 4} \rightarrow \bullet$. Three independent singlets can be made on the square, they can easily be obtained by diagonalizing the quadratic Casimir operator on four sites. The point group $C_{4v}$ naturally acts on this space and we can decompose the states in term of its irreducible representations: two singlets have ${\cal A}_1$ symmetry and the last one has ${\cal B}_2$ symmetry.

Diagonalizing ${\cal H}(\theta)$ on a 4-site plaquette in this restricted subspace leads to 3 eigenvalues $8 \sin (\theta )-6 \cos (\theta )\pm \sqrt{-76 \sin (2 \theta )-13 \cos (2 \theta )+85} / \sqrt{2}$ and $11 \sin (\theta )-8 \cos (\theta )$. Covering the lattice with such plaquettes and taking into account that uncorrelated bonds contribute to the total energy per site according to Eq.~({\ref{eq:uncorr}}) we obtain the variational energies :

\begin{align}
e_{\mathcal{P}_4}^{{\cal A}_1} &= -\frac{3}{2} \cos\theta + \frac{29}{12} \sin\theta  \nonumber \\ &\pm \frac{\sqrt{2}}{8} \sqrt{85 - 13 \cos(2\theta) -76 \sin(2 \theta)} \\
e_{\mathcal{P}_4}^{{\cal B}_2} &= -2\cos \theta + \frac{19}{6} \sin \theta
\end{align}

\paragraph{6-site plaquette state $\mathcal{P}_6$.}  For a single 6-site plaquette ($3\times 2$ rectangle), only one SU(6) singlet can be obtained if each site hosts the fundamental representation of SU(6). By construction, this state is antisymmetric with respect to any 2-site permutation. In the SU(4) language (obtained by a simple identification of the 6 states of the fundamental representation of SU(6) and the 6 states of the $\bf 6$ representation of SU(4)), this state can be viewed as a pairing of any pair of sites into the (antisymmetric) $\bf 15$ representation of SU(4). As a consequence, it becomes obvious that this is an eigenstate of ${\cal H}(\theta)$ with the energy $7\left ( -\cos \theta +(1/4) \sin \theta\right )$. As in the 4-site plaquette states, the variational energy of the uniform covering of the lattice with such 6-site plaquettes is given by

\begin{equation}
e_{\mathcal{P}_6} = -\frac{7}{6} \cos \theta + \frac{23}{36} \sin \theta.
\end{equation}

We only plot the ferromagnetic state, the lowest 4-site plaquette state and the 6-site plaquette state energies as a function of  $\theta$.

\section{Additional information for Exact Diagonalization}
\label{app:ed}

For exact diagonalization, we have computed the ground-state and low-energy excitations using a Lanczos algorithm on several finite-size clusters of $N_s$ sites with periodic boundary conditions, see Tab.~\ref{table:clusters}. Since we are mostly looking for a quantum spin liquid state, we have considered clusters that can accomodate the columnar phase (i.e. possess $(\pi,0)$ and $(0,\pi)$ momenta in their Brillouin zone) or not. Moreover, we have also considered one cluster ($N_s=12$) which is not a perfect square since its unit vectors are not perpendicular, which is not an issue for disordered phase~\cite{Chen2019}. Last, in order to reduce the size of the Hilbert space, we have used all space symmetries (translation and point-group) as well as the 3 Cartan U(1) symmetries (color conservation).

\begin{table}[!h]
\begin{tabular}{|c | c | c | c | c |}
\hline
cluster & ${\bf t}_1$ & ${\bf t}_2$ & can host a columnar phase & point group  \\
\hline
$8$     & $(2,2)$     & $(-2,2)$    & yes       & $C_{4v}$           \\
$10$     & $(1,3)$     & $(3,-1)$     & no         & $C_{4}$           \\
$12$     & $(1,3)$     & $(4,0)$      & no         & $C_{2}$           \\
$16$     & $(4,0)$     & $(0,4)$    & yes         & $C_{4v}$           \\
\hline
\end{tabular}
\caption{List of clusters studied with ED: number of sites $N_s$, the two unit vectors, compatibility with a dimerized phase,  and point-group symmetry.}
\label{table:clusters}
\end{table}

\section{PEPS with higher $\bf 6 6$ SU(6) symmetry}
 \label{subsec:tensors66}
 	
We now focus on the SU(6) $\textbf{66}$ symmetric point at 
$\theta=\pi/4$ (for which the GS is a non degenerate SU(6) singlet on finite clusters) and extend the tensor contruction in such a way that the PEPS now inherits the enlarged symmetry.
If we try to start from the previous SU(6) virtual space ${\cal V}_6=\bf 6\oplus {\bar 6}\oplus 1$, we are now left with tensors whose occupation numbers should be restricted to $n_{\rm occ}=\{1,0,3\}$ or $n_{\rm occ}=\{2,1,1\}$ on {\it both} A and B sites. It is easy to check that it is {\it impossible} to pave the square lattice with such tensors assuming the pairs of virtual states on the bonds are contracted
into $\bullet\bullet$ or $\bf 6\bar 6$ singlets. The same conclusion holds for ${\cal V}_6={\bf 6}\oplus {\bf\bar 6}\oplus {\bf 20}\oplus {\bf 1}$. We are there forced to introduce new/extra virtual degrees of freedom. The simplest choice of the SU(6) virtual space is
\begin{equation}
{\cal V}_6=\yng(1)\oplus\yng(1,1,1,1,1)\oplus\yng(1,1)\oplus\yng(1,1,1,1)\oplus\bullet\, \equiv \,{\bf 6}\oplus {\bf\bar 6}\oplus{\bf  15}\oplus\overline{\bf 15}\oplus{\bf 1} \, , 
\label{Eq:6615151a}  
\end{equation}
of dimension $D=43$. The corresponding PEPS tensors represent all the fusion channels of four of the five species  of (\ref{Eq:6615151a}) onto the physical state $\bf 6$. Mapping to SU(4) would require a virtual space with irreps of the same dimensions,
\begin{eqnarray}
\cal{V}&=&\yng(1,1)\oplus\yng(1,1)\oplus\yng(2,1,1)\oplus\yng(2,1,1)\oplus\bullet \nonumber \\ 
&& \nonumber\\
&\equiv& \,{\bf 6}\oplus {\bf 6}^*\oplus {\bf  15}\oplus{\bf 15}^* \oplus{\bf 1} \, , 
\label{Eq:6615151b}  
\end{eqnarray}
where both the $\bf 6$ (self-conjugate) and $\bf 15$ (adjoint) irreps occurs with multiplicity 2, ie with two ``colors''. Again, we use a $^*$ to distinguish the two copies. Unfortunately such a large $D=43$ bond dimension is untractable.

\begin{figure}
	\centering
	\includegraphics[width=0.5\textwidth]{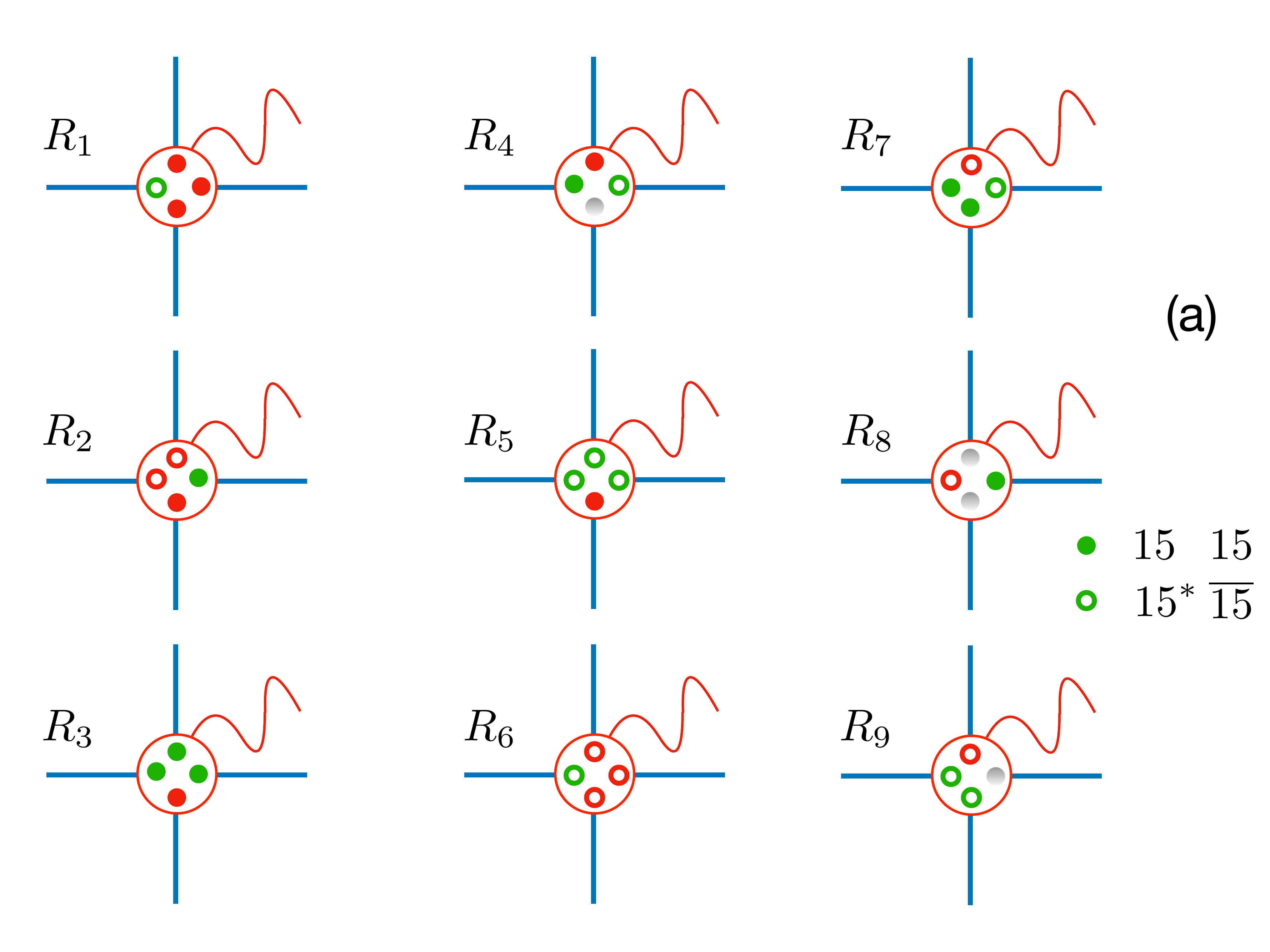}
	\vskip -0.2cm
		\includegraphics[width=0.5\textwidth]{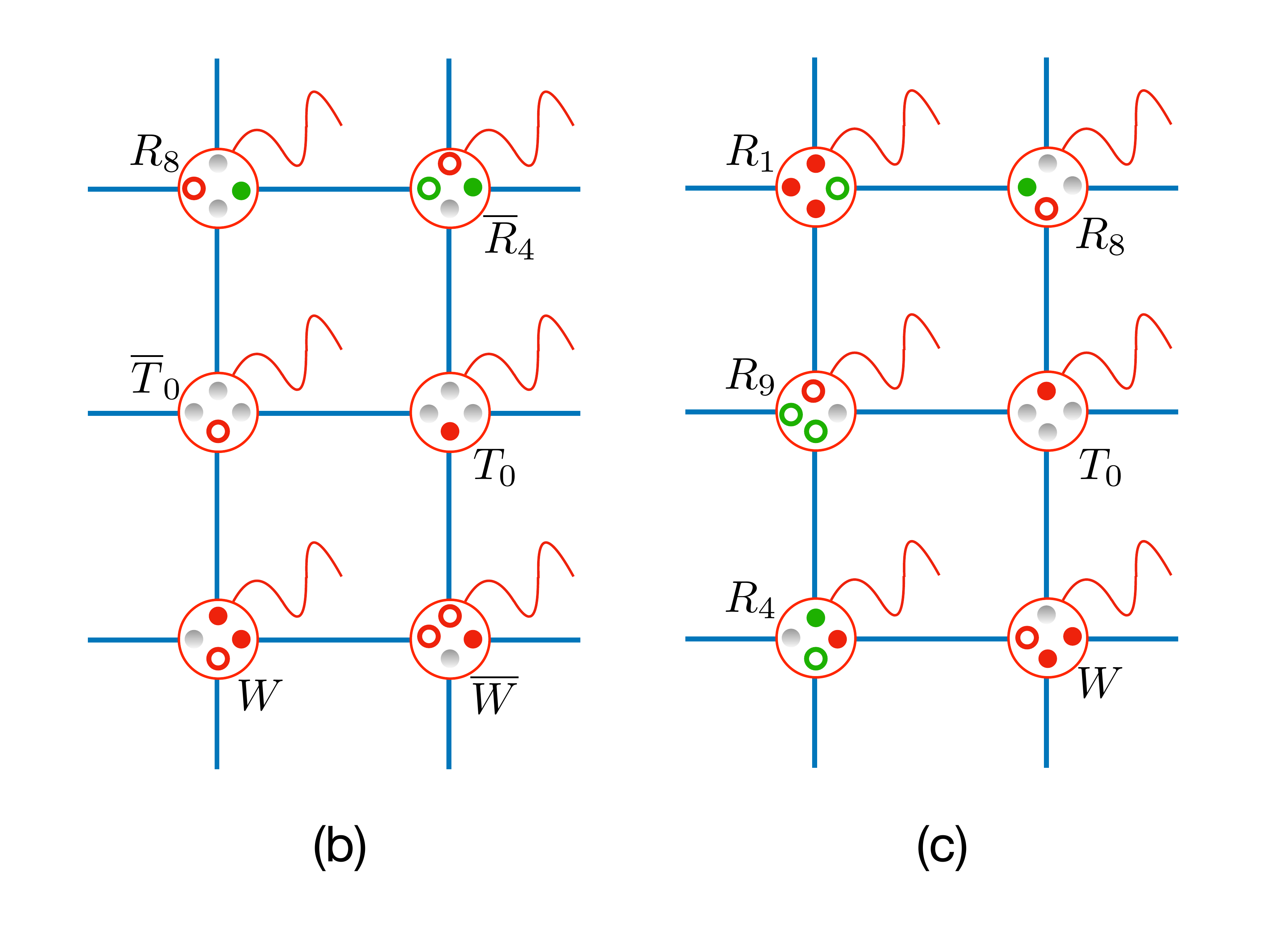}
			\caption{(a) The nine $R_i$ classes of $D=43$  tensors that can accomodate SU(6) fusion rules -- see Table~\ref{tab:tens_occ}. Color-conjugated tensors (same tensors) have to be used on the A and B sites to accomodate the $\bf 6 \bar 6$ ($\bf 6 6$) SU(6) symmetry as shown in (b) and (c), respectively.}
	\label{fig:TN2b}
\end{figure}

 In order to accomodate SU(6) fusion channels, we are left with only eleven classes defined by their possible occupations $n_{\rm occ}$ of the five virtual particles, as shown in Table~\ref{tab:tens_occ}.
The corresponding (classes of) tensors are drawn in Fig.~\ref{fig:TN2b}(a). Possible configurations of the PEPS after contracting some of these tensors is shown on Figs.~\ref{fig:TN2b}(b) and \ref{fig:TN2b}(c). The full tensor classification of the D=43 virtual space is given in Table \ref{tab:tens_occ3}.

Note that the C-symmetry corresponding to charge conjugation in the SU(6) case (and acts on the physical space as well) maps, in the SU(4) case, to a {\it gauge} symmetry $\cal C$ defined as color exchange acting only on the virtual space. Since the C symmetry is explicitly broken in the SU(6) PEPS (by construction, enforcing the physical irrep $\bf 6$ on every site), its related gauge symmetry has to be broken in a very specific way in the corresponding SU(4) PEPS, as e.g.\ shown in Table~\ref{tab:tens_occ}. In fact, the gauge ${\cal C}$-symmetry can be decomposed as ${\cal C}={\cal C}_1 {\cal C}_2$, where ${\cal C}_1$ and ${\cal C}_2$ color-exchange only the two $\bf 6$ irreps or the two $\bf 15$ irreps, respectively. These gauge transformations belong, in fact, to a larger $\mathrm{SU}(2)\times \mathrm{SU}(2)$ gauge group. 
Hence three other classes of tensors 
%$W_i^\prime$, $W_i^{\prime\prime}$ and $\overline{W_i}$ 
can be easily obtained from the $R_i$ classes by applying ${\cal C}_1$ only, ${\cal C}_2$ only or their product ${\cal C}$, respectively (see Table (\ref{tab:tens_occ})). 
At the SU(6)-symmetric points, any one -- but only one -- of the four classes can be used to construct a SU(6) singlet. Note that for $\bf 6\bar6$ SU(6) symmetry, the previous $\bf 6\oplus 6\oplus 1$ PEPS can be extended using the eleven classes of $D=43$ tensors on the A sites
and their color-conjugate both in $\bf 6$ and $\bf 15$ of the B sites, as shown in Fig.~\ref{fig:TN2b}(c). 

\begin{table}
	\vskip 0.5truecm
	\centering
	\begin{tabular}{CCCC}
		\hline
		\hline
		\mathrm{Class} & D & {\cal V} & \quad n_\mathrm{occ\, [D=43]}\quad \\
		\hline
		T_0& \quad 7 \quad& \textbf{6}\oplus\textbf{1} & \quad \{1,0,0,0,3\} \quad \\
		\hline
		W& \quad 13\quad & \; \textbf{6}\oplus{\textbf{6}}\oplus\textbf{1} \; & \quad {\{2,1,0,0,1\}} \quad \\
		\hline
		R_1&\quad 43 \quad& \quad \textbf{6}\oplus{\textbf{6}}\oplus{\textbf{15}}\oplus{\textbf{15}}\oplus\textbf{1} \quad &\quad  {  \{3,0,0,1,0\}} \quad   \\
		R_2&\quad 43 \quad& \quad \textbf{6}\oplus{\textbf{6}}\oplus{\textbf{15}}\oplus{\textbf{15}}\oplus\textbf{1} \quad &\quad  {  \{1,2,1,0,0\}} \quad   \\
		R_3&\quad 43 \quad& \quad \textbf{6}\oplus{\textbf{6}}\oplus{\textbf{15}}\oplus{\textbf{15}}\oplus\textbf{1} \quad &\quad  {  \{1,0,3,0,0\}} \quad   \\
		R_4&\quad 43 \quad& \quad \textbf{6}\oplus{\textbf{6}}\oplus{\textbf{15}}\oplus{\textbf{15}}\oplus\textbf{1} \quad &\quad  {  \{1,0,1,1,1\}} \quad   \\
		R_5&\quad 43 \quad& \quad \textbf{6}\oplus{\textbf{6}}\oplus{\textbf{15}}\oplus{\textbf{15}}\oplus\textbf{1} \quad &\quad  {  \{1,0,0,3,0\}} \quad   \\
		R_6&\quad 43 \quad& \quad \textbf{6}\oplus{\textbf{6}}\oplus{\textbf{15}}\oplus{\textbf{15}}\oplus\textbf{1} \quad &\quad  {  \{0,3,0,1,0\}} \quad   \\
		R_7&\quad 43 \quad& \quad \textbf{6}\oplus{\textbf{6}}\oplus{\textbf{15}}\oplus{\textbf{15}}\oplus\textbf{1} \quad &\quad  {  \{0,1,2,1,0\}} \quad   \\
		R_8&\quad 43 \quad& \quad \textbf{6}\oplus{\textbf{6}}\oplus{\textbf{15}}\oplus{\textbf{15}}\oplus\textbf{1} \quad &\quad  {  \{0,1,1,0,2\}} \quad   \\
		R_9&\quad 43 \quad& \quad \textbf{6}\oplus{\textbf{6}}\oplus{\textbf{15}}\oplus{\textbf{15}}\oplus\textbf{1} \quad &\quad  {  \{0,1,0,2,1\}} \quad   \\
		\hline
		\hline
	\end{tabular}
	\caption{Eleven classes of SU(4)-symmetric tensors which can accomodate SU($6$)-symmetry, classified according to their virtual space and occupation numbers in the $D=43$ largest SU(4) virtual space. 
	}
	\label{tab:tens_occ}
\end{table}

\begin{table}
	\centering
	\begin{tabular}{CCC}
		\hline
		\hline
%		\begin{align*}\begin{array}{ccc}
\text{Occupation number} & \text{SU(4)} & \text{SU(6)} \\
{\color{blue} \{0,1,0,0,3\} }& {\color{blue} 4 }& {\color{blue} 0} \\
\{0,1,0,1,2\} & 12 & 0 \\
\{0,1,0,2,1\} & 48 & 12 \\
\{0,1,0,3,0\} & 64 & 0 \\
\{0,1,1,0,2\} & 12 & 12 \\
\{0,1,1,1,1\} & 96 & 0 \\
\{0,1,1,2,0\} & 192 & 0 \\
\{0,1,2,0,1\} & 48 & 0 \\
\{0,1,2,1,0\} & 192 & 48 \\
\{0,1,3,0,0\} & 64 & 0 \\
{\color{blue} \{0,3,0,0,1\}} &{\color{blue}  12} & {\color{blue} 0} \\
\{0,3,0,1,0\} & 28 & 4 \\
\{0,3,1,0,0\} & 28 & 0 \\
{\color{red}  \{1,0,0,0,3\}} & {\color{red} 4} & {\color{red} 4} \\
\{1,0,0,1,2\} & 12 & 0 \\
\{1,0,0,2,1\} & 48 & 0 \\
\{1,0,0,3,0\} & 64 & 12 \\
\{1,0,1,0,2\} & 12 & 0 \\
\{1,0,1,1,1\} & 96 & 48 \\
\{1,0,1,2,0\} & 192 & 0 \\
\{1,0,2,0,1\} & 48 & 0 \\
\{1,0,2,1,0\} & 192 & 0 \\
\{1,0,3,0,0\} & 64 & 12 \\
{\color{blue} \{1,2,0,0,1\} }& {\color{blue} 36} & {\color{blue} 0} \\
\{1,2,0,1,0\} & 84 & 0 \\
\{1,2,1,0,0\} & 84 & 36 \\
{\color{blue} \{2,1,0,0,1\} }& {\color{blue} 36 }& {\color{blue} 24} \\
\{2,1,0,1,0\} & 84 & 0 \\
\{2,1,1,0,0\} & 84 & 0 \\
{\color{blue} \{3,0,0,0,1\} }& {\color{blue} 12} & {\color{blue} 0} \\
\{3,0,0,1,0\} & 28 & 12 \\
\{3,0,1,0,0\} & 28 & 0 \\
%\end{array}\end{align*}
		\hline
\hline
\end{tabular}
\label{tab:tens_occ3}
\caption{Full list of the classes of $D=43$ ($\bf 6\oplus 6\oplus 15\oplus 15\oplus1$) SU(4) symmetric tensors in terms of the occupation numbers of the five virtual particles. The second (third) column shows the number of SU(4) (SU(6)) fusion channels -- i.e.\ the total number of $\mathrm{SU}(N)$ symmetric tensors (that could be further classified in terms of the irreps of the $C_{4v}$ point group, ${\cal A}_1$, ${\cal A}_2$, ${\cal B}_1$, ${\cal B}_2$ and ${\cal E}$). The $\bf 6\oplus 1$ $T_0$ tensor is marked in red.  The $\bf 6\oplus 6\oplus 1$ tensors are marked in blue.  Note that the total occupancy of the $\bf 6$ particles is odd, reflecting a gauge $\mathbb{Z}_2$ symmetry.}
\end{table}

\section{Gauge symmetries}
\label{subsec:gauge}

The gauge symmetries of each of the tensor families can be viewed as charge conservation in the fusion process of the four vitual particles into the physical one. Therefore, one has to assign an integer charge to both virtual and physical degrees of freedom. It is natural to expect $\mathbb{Z}_4$ gauge symmetry for SU(4)-symmetric tensors in which case one should consider a minimal virtual space ${\cal V}=\textbf{4}\oplus\overline{\textbf{4}}\oplus\textbf{6}\oplus \textbf{1}$ to encode all particle types with charges $q=\{1,-1,2,0\}$ (mod 4) and the onsite charge conservation reads $q\cdot n_{\rm occ}=2$ (mod 4). However, it is easy to check that, for the virtual spaces that do not contain all four SU(4) charges, the gauge symmetry may be reduced to $\mathbb{Z}_2$, as shown in Table~\ref{tab:gauge}.  For SU(6) symmetry, the $D=63$ ${\cal V}_6=\bf 6\oplus\overline{6}\oplus 15\oplus \overline{15} \oplus 20\oplus 1$ tensors contains four of the six elementary virtual particles of charges $q_6=\{1,-1,2,-2,3,0\}$ (mod 6) and the SU(6) conservation rules state that $q_6\cdot n_{\rm occ}=1$ (mod 6), providing the desired $\mathbb{Z}_6$ gauge symmetry. The gauge symmetry is also present for smaller bond dimension, except for ${\cal V}_6=\bf 6\oplus 1$, for which it is reduced to $\mathbb{Z}_2$, as shown in Table~\ref{tab:gauge}.

\begin{table}
	\vskip 0.5truecm
	\centering
	\begin{tabular}{CCCCCC}
		\hline
		\hline
		\\[-9pt]
		D & {\cal V} & SU(4) / SU(6) & \textbf{6}\overline{\textbf{6}} & \textbf{66} \\
		\hline
		7& \textbf{6}\oplus\textbf{1} & \mathbb{Z}_2 / \mathbb{Z}_2 & \sym& \nsym\\
		8& \textbf{4}\oplus\overline{\textbf{4}} & \mathbb{Z}_2 / - & -&  - \\	
		9& \textbf{4}\oplus\overline{\textbf{4}}\oplus \textbf{1} & \mathbb{Z}_4 / - & -&  - \\	
		13 & \; \textbf{6}\oplus{\textbf{6}}\oplus\textbf{1} \; & \mathbb{Z}_2 /   \mathbb{Z}_6   &\sym &  \nsym\\
		15& \textbf{4}\oplus\overline{\textbf{4}}\oplus\textbf{6}\oplus \textbf{1} & \mathbb{Z}_4  / - &-&-\\	
		21 & \quad \textbf{4}\oplus\overline{\textbf{4}}\oplus \textbf{6}\oplus{\textbf{6}} \oplus\textbf{1} \quad &  \mathbb{Z}_4 /  -  & -& - \\
		33 & \quad \textbf{6}\oplus{\textbf{6}}\oplus{\textbf{20}}\oplus\textbf{1} \quad &  \mathbb{Z}_2 /   \mathbb{Z}_6  &\sym &\nsym  \\
		43 & \quad \textbf{6}\oplus{\textbf{6}}\oplus{\textbf{15}}\oplus{\textbf{15}}\oplus\textbf{1} \quad &  \mathbb{Z}_2 /  \mathbb{Z}_6  &\sym & \sym  \\
		63 & \quad \textbf{6}\oplus{\textbf{6}}\oplus{\textbf{15}}\oplus{\textbf{15}}\oplus \textbf{20} \oplus\textbf{1} \quad &  \mathbb{Z}_2 /  \mathbb{Z}_6  &\sym & \sym\\
		\hline		\hline
	\end{tabular}
	\caption{ Gauge  symmetries of the various SU(4)-symmetric tensor families and of their imbeded SU(6)-symmetric sub-families, if they exist. The last two columns indicate whether $(\sym)$ or not  ($\nsym$) the SU(6) symmetric tensors can describe the $\bf 6\overline{6}$ and $\bf 66$ symmetric points, respectively.
	}
	\label{tab:gauge}
\end{table}

%%%%%%%%%%%%%%%%%%%%%%%%%%%%%%%%%%%%%%%%%%%%%%%%%%%%%%
\section{Tensor expressions}\label{app:tensors}
The expression of the tensors $T_i$ can be found in the supplemental material of \cite{gauthe_su4_2019}. We provide here the coefficients of the unormalized tensors $X_{31}$ (Table \ref{Table:X31}) and $W_1$ (Table \ref{Table:W1}), which are integer values. The first index labels the physical variable (varying from 0 to 5), with weights $(1, 0, -1)$, $(1, -1, 1)$, $(0, -1, 0)$, $(0, 1, 0)$, $(-1, 1, -1)$ and $(-1, 0, 1)$, respectively. The four subsequent indices the virtual variables on the links (in e.g.\ clockwise direction). For $X_{31}$, this corresponds to the virtual space $\mathcal{V}=\textbf{4}\oplus \overline{\textbf{4}}$
with weights $(1, 0, 0)$, $(-1, 1, 0)$, $(0, -1, 1)$, $(0, 0, -1)$, $(-1, 0, 0)$, $(1, -1, 0)$, $(0, 1, -1)$ and $(0, 0, 1)$. For $W_1$, this corresponds to the SU(6) virtual space $\mathcal{V}=\textbf{6}\oplus \overline{\textbf{6}}\oplus \textbf{1}$, with weights $(1,0,0,0,0)$, $(-1,1,0,0,0)$, $(0,-1,1,0,0)$, $(0,0,-1,1,0)$, $(0,0,0,-1,1)$, $(0,0,0,0,-1)$, $(-1,0,0,0,0)$, $(1,-1,0,0,0)$, $(0,1,-1,0,0)$, $(0,0,1,-1,0)$, $(0,0,0,1,-1)$, $(0,0,0,0,1)$ and $(0,0,0,0,0)$.

\begin{longtable}{MMM}
\hline \hline
X_{31}[0,0,0,3,4] =  1  &  X_{31}[0,0,0,4,3] =  1  &  X_{31}[0,0,1,5,3] =  1  \\
X_{31}[0,0,2,6,3] =  1  &  X_{31}[0,0,3,0,4] = -2  &  X_{31}[0,0,3,1,5] = -1  \\
X_{31}[0,0,3,2,6] = -1  &  X_{31}[0,0,3,3,7] = -1  &  X_{31}[0,0,3,4,0] =  1  \\
X_{31}[0,0,3,5,1] =  1  &  X_{31}[0,0,3,6,2] =  1  &  X_{31}[0,0,3,7,3] =  2  \\
X_{31}[0,0,4,0,3] = -2  &  X_{31}[0,0,4,3,0] =  1  &  X_{31}[0,0,5,1,3] = -1  \\
X_{31}[0,0,6,2,3] = -1  &  X_{31}[0,0,7,3,3] = -1  &  X_{31}[0,1,0,3,5] =  1  \\
X_{31}[0,1,3,0,5] = -1  &  X_{31}[0,1,5,0,3] = -1  &  X_{31}[0,1,5,3,0] =  1  \\
X_{31}[0,2,0,3,6] =  1  &  X_{31}[0,2,3,0,6] = -1  &  X_{31}[0,2,6,0,3] = -1  \\
X_{31}[0,2,6,3,0] =  1  &  X_{31}[0,3,0,0,4] =  1  &  X_{31}[0,3,0,1,5] =  1  \\
X_{31}[0,3,0,2,6] =  1  &  X_{31}[0,3,0,3,7] =  2  &  X_{31}[0,3,0,4,0] = -2  \\
X_{31}[0,3,0,5,1] = -1  &  X_{31}[0,3,0,6,2] = -1  &  X_{31}[0,3,0,7,3] = -1  \\
X_{31}[0,3,1,5,0] = -1  &  X_{31}[0,3,2,6,0] = -1  &  X_{31}[0,3,3,0,7] = -1  \\
X_{31}[0,3,3,7,0] = -1  &  X_{31}[0,3,4,0,0] =  1  &  X_{31}[0,3,5,1,0] =  1  \\
X_{31}[0,3,6,2,0] =  1  &  X_{31}[0,3,7,0,3] = -1  &  X_{31}[0,3,7,3,0] =  2  \\
X_{31}[0,4,0,0,3] =  1  &  X_{31}[0,4,0,3,0] = -2  &  X_{31}[0,4,3,0,0] =  1  \\
X_{31}[0,5,0,3,1] = -1  &  X_{31}[0,5,1,0,3] =  1  &  X_{31}[0,5,1,3,0] = -1  \\
X_{31}[0,5,3,0,1] =  1  &  X_{31}[0,6,0,3,2] = -1  &  X_{31}[0,6,2,0,3] =  1  \\
X_{31}[0,6,2,3,0] = -1  &  X_{31}[0,6,3,0,2] =  1  &  X_{31}[0,7,0,3,3] = -1  \\
X_{31}[0,7,3,0,3] =  2  &  X_{31}[0,7,3,3,0] = -1  &  X_{31}[1,0,0,2,4] =  1  \\
X_{31}[1,0,0,4,2] =  1  &  X_{31}[1,0,1,5,2] =  1  &  X_{31}[1,0,2,0,4] = -2  \\
X_{31}[1,0,2,1,5] = -1  &  X_{31}[1,0,2,2,6] = -1  &  X_{31}[1,0,2,3,7] = -1  \\
X_{31}[1,0,2,4,0] =  1  &  X_{31}[1,0,2,5,1] =  1  &  X_{31}[1,0,2,6,2] =  2  \\
X_{31}[1,0,2,7,3] =  1  &  X_{31}[1,0,3,7,2] =  1  &  X_{31}[1,0,4,0,2] = -2  \\
X_{31}[1,0,4,2,0] =  1  &  X_{31}[1,0,5,1,2] = -1  &  X_{31}[1,0,6,2,2] = -1  \\
X_{31}[1,0,7,3,2] = -1  &  X_{31}[1,1,0,2,5] =  1  &  X_{31}[1,1,2,0,5] = -1  \\
X_{31}[1,1,5,0,2] = -1  &  X_{31}[1,1,5,2,0] =  1  &  X_{31}[1,2,0,0,4] =  1  \\
X_{31}[1,2,0,1,5] =  1  &  X_{31}[1,2,0,2,6] =  2  &  X_{31}[1,2,0,3,7] =  1  \\
X_{31}[1,2,0,4,0] = -2  &  X_{31}[1,2,0,5,1] = -1  &  X_{31}[1,2,0,6,2] = -1  \\
X_{31}[1,2,0,7,3] = -1  &  X_{31}[1,2,1,5,0] = -1  &  X_{31}[1,2,2,0,6] = -1  \\
X_{31}[1,2,2,6,0] = -1  &  X_{31}[1,2,3,7,0] = -1  &  X_{31}[1,2,4,0,0] =  1  \\
X_{31}[1,2,5,1,0] =  1  &  X_{31}[1,2,6,0,2] = -1  &  X_{31}[1,2,6,2,0] =  2  \\
X_{31}[1,2,7,3,0] =  1  &  X_{31}[1,3,0,2,7] =  1  &  X_{31}[1,3,2,0,7] = -1  \\
X_{31}[1,3,7,0,2] = -1  &  X_{31}[1,3,7,2,0] =  1  &  X_{31}[1,4,0,0,2] =  1  \\
X_{31}[1,4,0,2,0] = -2  &  X_{31}[1,4,2,0,0] =  1  &  X_{31}[1,5,0,2,1] = -1  \\
X_{31}[1,5,1,0,2] =  1  &  X_{31}[1,5,1,2,0] = -1  &  X_{31}[1,5,2,0,1] =  1  \\
X_{31}[1,6,0,2,2] = -1  &  X_{31}[1,6,2,0,2] =  2  &  X_{31}[1,6,2,2,0] = -1  \\
X_{31}[1,7,0,2,3] = -1  &  X_{31}[1,7,2,0,3] =  1  &  X_{31}[1,7,3,0,2] =  1  \\
X_{31}[1,7,3,2,0] = -1  &  X_{31}[2,0,2,3,4] =  1  &  X_{31}[2,0,3,2,4] = -1  \\
X_{31}[2,0,4,2,3] = -1  &  X_{31}[2,0,4,3,2] =  1  &  X_{31}[2,1,2,3,5] =  1  \\
X_{31}[2,1,3,2,5] = -1  &  X_{31}[2,1,5,2,3] = -1  &  X_{31}[2,1,5,3,2] =  1  \\
X_{31}[2,2,0,4,3] =  1  &  X_{31}[2,2,1,5,3] =  1  &  X_{31}[2,2,2,3,6] =  1  \\
X_{31}[2,2,2,6,3] =  1  &  X_{31}[2,2,3,0,4] = -1  &  X_{31}[2,2,3,1,5] = -1  \\
X_{31}[2,2,3,2,6] = -2  &  X_{31}[2,2,3,3,7] = -1  &  X_{31}[2,2,3,4,0] =  1  \\
X_{31}[2,2,3,5,1] =  1  &  X_{31}[2,2,3,6,2] =  1  &  X_{31}[2,2,3,7,3] =  2  \\
X_{31}[2,2,4,0,3] = -1  &  X_{31}[2,2,5,1,3] = -1  &  X_{31}[2,2,6,2,3] = -2  \\
X_{31}[2,2,6,3,2] =  1  &  X_{31}[2,2,7,3,3] = -1  &  X_{31}[2,3,0,4,2] = -1  \\
X_{31}[2,3,1,5,2] = -1  &  X_{31}[2,3,2,0,4] =  1  &  X_{31}[2,3,2,1,5] =  1  \\
X_{31}[2,3,2,2,6] =  1  &  X_{31}[2,3,2,3,7] =  2  &  X_{31}[2,3,2,4,0] = -1  \\
X_{31}[2,3,2,5,1] = -1  &  X_{31}[2,3,2,6,2] = -2  &  X_{31}[2,3,2,7,3] = -1  \\
X_{31}[2,3,3,2,7] = -1  &  X_{31}[2,3,3,7,2] = -1  &  X_{31}[2,3,4,0,2] =  1  \\
X_{31}[2,3,5,1,2] =  1  &  X_{31}[2,3,6,2,2] =  1  &  X_{31}[2,3,7,2,3] = -1  \\
X_{31}[2,3,7,3,2] =  2  &  X_{31}[2,4,0,2,3] =  1  &  X_{31}[2,4,0,3,2] = -1  \\
X_{31}[2,4,2,3,0] = -1  &  X_{31}[2,4,3,2,0] =  1  &  X_{31}[2,5,1,2,3] =  1  \\
X_{31}[2,5,1,3,2] = -1  &  X_{31}[2,5,2,3,1] = -1  &  X_{31}[2,5,3,2,1] =  1  \\
X_{31}[2,6,2,2,3] =  1  &  X_{31}[2,6,2,3,2] = -2  &  X_{31}[2,6,3,2,2] =  1  \\
X_{31}[2,7,2,3,3] = -1  &  X_{31}[2,7,3,2,3] =  2  &  X_{31}[2,7,3,3,2] = -1  \\
X_{31}[3,0,0,1,4] =  1  &  X_{31}[3,0,0,4,1] =  1  &  X_{31}[3,0,1,0,4] = -2  \\
X_{31}[3,0,1,1,5] = -1  &  X_{31}[3,0,1,2,6] = -1  &  X_{31}[3,0,1,3,7] = -1  \\
X_{31}[3,0,1,4,0] =  1  &  X_{31}[3,0,1,5,1] =  2  &  X_{31}[3,0,1,6,2] =  1  \\
X_{31}[3,0,1,7,3] =  1  &  X_{31}[3,0,2,6,1] =  1  &  X_{31}[3,0,3,7,1] =  1  \\
X_{31}[3,0,4,0,1] = -2  &  X_{31}[3,0,4,1,0] =  1  &  X_{31}[3,0,5,1,1] = -1  \\
X_{31}[3,0,6,2,1] = -1  &  X_{31}[3,0,7,3,1] = -1  &  X_{31}[3,1,0,0,4] =  1  \\
X_{31}[3,1,0,1,5] =  2  &  X_{31}[3,1,0,2,6] =  1  &  X_{31}[3,1,0,3,7] =  1  \\
X_{31}[3,1,0,4,0] = -2  &  X_{31}[3,1,0,5,1] = -1  &  X_{31}[3,1,0,6,2] = -1  \\
X_{31}[3,1,0,7,3] = -1  &  X_{31}[3,1,1,0,5] = -1  &  X_{31}[3,1,1,5,0] = -1  \\
X_{31}[3,1,2,6,0] = -1  &  X_{31}[3,1,3,7,0] = -1  &  X_{31}[3,1,4,0,0] =  1  \\
X_{31}[3,1,5,0,1] = -1  &  X_{31}[3,1,5,1,0] =  2  &  X_{31}[3,1,6,2,0] =  1  \\
X_{31}[3,1,7,3,0] =  1  &  X_{31}[3,2,0,1,6] =  1  &  X_{31}[3,2,1,0,6] = -1  \\
X_{31}[3,2,6,0,1] = -1  &  X_{31}[3,2,6,1,0] =  1  &  X_{31}[3,3,0,1,7] =  1  \\
X_{31}[3,3,1,0,7] = -1  &  X_{31}[3,3,7,0,1] = -1  &  X_{31}[3,3,7,1,0] =  1  \\
X_{31}[3,4,0,0,1] =  1  &  X_{31}[3,4,0,1,0] = -2  &  X_{31}[3,4,1,0,0] =  1  \\
X_{31}[3,5,0,1,1] = -1  &  X_{31}[3,5,1,0,1] =  2  &  X_{31}[3,5,1,1,0] = -1  \\
X_{31}[3,6,0,1,2] = -1  &  X_{31}[3,6,1,0,2] =  1  &  X_{31}[3,6,2,0,1] =  1  \\
X_{31}[3,6,2,1,0] = -1  &  X_{31}[3,7,0,1,3] = -1  &  X_{31}[3,7,1,0,3] =  1  \\
X_{31}[3,7,3,0,1] =  1  &  X_{31}[3,7,3,1,0] = -1  &  X_{31}[4,0,1,3,4] =  1  \\
X_{31}[4,0,3,1,4] = -1  &  X_{31}[4,0,4,1,3] = -1  &  X_{31}[4,0,4,3,1] =  1  \\
X_{31}[4,1,0,4,3] =  1  &  X_{31}[4,1,1,3,5] =  1  &  X_{31}[4,1,1,5,3] =  1  \\
X_{31}[4,1,2,6,3] =  1  &  X_{31}[4,1,3,0,4] = -1  &  X_{31}[4,1,3,1,5] = -2  \\
X_{31}[4,1,3,2,6] = -1  &  X_{31}[4,1,3,3,7] = -1  &  X_{31}[4,1,3,4,0] =  1  \\
X_{31}[4,1,3,5,1] =  1  &  X_{31}[4,1,3,6,2] =  1  &  X_{31}[4,1,3,7,3] =  2  \\
X_{31}[4,1,4,0,3] = -1  &  X_{31}[4,1,5,1,3] = -2  &  X_{31}[4,1,5,3,1] =  1  \\
X_{31}[4,1,6,2,3] = -1  &  X_{31}[4,1,7,3,3] = -1  &  X_{31}[4,2,1,3,6] =  1  \\
X_{31}[4,2,3,1,6] = -1  &  X_{31}[4,2,6,1,3] = -1  &  X_{31}[4,2,6,3,1] =  1  \\
X_{31}[4,3,0,4,1] = -1  &  X_{31}[4,3,1,0,4] =  1  &  X_{31}[4,3,1,1,5] =  1  \\
X_{31}[4,3,1,2,6] =  1  &  X_{31}[4,3,1,3,7] =  2  &  X_{31}[4,3,1,4,0] = -1  \\
X_{31}[4,3,1,5,1] = -2  &  X_{31}[4,3,1,6,2] = -1  &  X_{31}[4,3,1,7,3] = -1  \\
X_{31}[4,3,2,6,1] = -1  &  X_{31}[4,3,3,1,7] = -1  &  X_{31}[4,3,3,7,1] = -1  \\
X_{31}[4,3,4,0,1] =  1  &  X_{31}[4,3,5,1,1] =  1  &  X_{31}[4,3,6,2,1] =  1  \\
X_{31}[4,3,7,1,3] = -1  &  X_{31}[4,3,7,3,1] =  2  &  X_{31}[4,4,0,1,3] =  1  \\
X_{31}[4,4,0,3,1] = -1  &  X_{31}[4,4,1,3,0] = -1  &  X_{31}[4,4,3,1,0] =  1  \\
X_{31}[4,5,1,1,3] =  1  &  X_{31}[4,5,1,3,1] = -2  &  X_{31}[4,5,3,1,1] =  1  \\
X_{31}[4,6,1,3,2] = -1  &  X_{31}[4,6,2,1,3] =  1  &  X_{31}[4,6,2,3,1] = -1  \\
X_{31}[4,6,3,1,2] =  1  &  X_{31}[4,7,1,3,3] = -1  &  X_{31}[4,7,3,1,3] =  2  \\
X_{31}[4,7,3,3,1] = -1  &  X_{31}[5,0,1,2,4] =  1  &  X_{31}[5,0,2,1,4] = -1  \\
X_{31}[5,0,4,1,2] = -1  &  X_{31}[5,0,4,2,1] =  1  &  X_{31}[5,1,0,4,2] =  1  \\
X_{31}[5,1,1,2,5] =  1  &  X_{31}[5,1,1,5,2] =  1  &  X_{31}[5,1,2,0,4] = -1  \\
X_{31}[5,1,2,1,5] = -2  &  X_{31}[5,1,2,2,6] = -1  &  X_{31}[5,1,2,3,7] = -1  \\
X_{31}[5,1,2,4,0] =  1  &  X_{31}[5,1,2,5,1] =  1  &  X_{31}[5,1,2,6,2] =  2  \\
X_{31}[5,1,2,7,3] =  1  &  X_{31}[5,1,3,7,2] =  1  &  X_{31}[5,1,4,0,2] = -1  \\
X_{31}[5,1,5,1,2] = -2  &  X_{31}[5,1,5,2,1] =  1  &  X_{31}[5,1,6,2,2] = -1  \\
X_{31}[5,1,7,3,2] = -1  &  X_{31}[5,2,0,4,1] = -1  &  X_{31}[5,2,1,0,4] =  1  \\
X_{31}[5,2,1,1,5] =  1  &  X_{31}[5,2,1,2,6] =  2  &  X_{31}[5,2,1,3,7] =  1  \\
X_{31}[5,2,1,4,0] = -1  &  X_{31}[5,2,1,5,1] = -2  &  X_{31}[5,2,1,6,2] = -1  \\
X_{31}[5,2,1,7,3] = -1  &  X_{31}[5,2,2,1,6] = -1  &  X_{31}[5,2,2,6,1] = -1  \\
X_{31}[5,2,3,7,1] = -1  &  X_{31}[5,2,4,0,1] =  1  &  X_{31}[5,2,5,1,1] =  1  \\
X_{31}[5,2,6,1,2] = -1  &  X_{31}[5,2,6,2,1] =  2  &  X_{31}[5,2,7,3,1] =  1  \\
X_{31}[5,3,1,2,7] =  1  &  X_{31}[5,3,2,1,7] = -1  &  X_{31}[5,3,7,1,2] = -1  \\
X_{31}[5,3,7,2,1] =  1  &  X_{31}[5,4,0,1,2] =  1  &  X_{31}[5,4,0,2,1] = -1  \\
X_{31}[5,4,1,2,0] = -1  &  X_{31}[5,4,2,1,0] =  1  &  X_{31}[5,5,1,1,2] =  1  \\
X_{31}[5,5,1,2,1] = -2  &  X_{31}[5,5,2,1,1] =  1  &  X_{31}[5,6,1,2,2] = -1  \\
X_{31}[5,6,2,1,2] =  2  &  X_{31}[5,6,2,2,1] = -1  &  X_{31}[5,7,1,2,3] = -1  \\
X_{31}[5,7,2,1,3] =  1  &  X_{31}[5,7,3,1,2] =  1  &  X_{31}[5,7,3,2,1] = -1  \\

\hline \hline
\caption{Tensor $X_{31}$ (multiplied by $4\sqrt{30}$).}
\label{Table:X31}
\end{longtable}\begin{longtable}{MMM}
\hline \hline
W_1[0,0,0,6,12] =  2  &  W_1[0,0,0,12,6] =  2  &  W_1[0,0,1,7,12] =  1  \\
W_1[0,0,1,12,7] =  1  &  W_1[0,0,2,8,12] =  1  &  W_1[0,0,2,12,8] =  1  \\
W_1[0,0,3,9,12] =  1  &  W_1[0,0,3,12,9] =  1  &  W_1[0,0,4,10,12] =  1  \\
W_1[0,0,4,12,10] =  1  &  W_1[0,0,5,11,12] =  1  &  W_1[0,0,5,12,11] =  1  \\
W_1[0,0,6,0,12] =  2  &  W_1[0,0,6,12,0] =  2  &  W_1[0,0,7,1,12] =  1  \\
W_1[0,0,7,12,1] =  1  &  W_1[0,0,8,2,12] =  1  &  W_1[0,0,8,12,2] =  1  \\
W_1[0,0,9,3,12] =  1  &  W_1[0,0,9,12,3] =  1  &  W_1[0,0,10,4,12] =  1  \\
W_1[0,0,10,12,4] =  1  &  W_1[0,0,11,5,12] =  1  &  W_1[0,0,11,12,5] =  1  \\
W_1[0,0,12,0,6] =  2  &  W_1[0,0,12,1,7] =  1  &  W_1[0,0,12,2,8] =  1  \\
W_1[0,0,12,3,9] =  1  &  W_1[0,0,12,4,10] =  1  &  W_1[0,0,12,5,11] =  1  \\
W_1[0,0,12,6,0] =  2  &  W_1[0,0,12,7,1] =  1  &  W_1[0,0,12,8,2] =  1  \\
W_1[0,0,12,9,3] =  1  &  W_1[0,0,12,10,4] =  1  &  W_1[0,0,12,11,5] =  1  \\
W_1[0,1,0,7,12] =  1  &  W_1[0,1,0,12,7] =  1  &  W_1[0,1,7,0,12] =  1  \\
W_1[0,1,7,12,0] =  1  &  W_1[0,1,12,0,7] =  1  &  W_1[0,1,12,7,0] =  1  \\
W_1[0,2,0,8,12] =  1  &  W_1[0,2,0,12,8] =  1  &  W_1[0,2,8,0,12] =  1  \\
W_1[0,2,8,12,0] =  1  &  W_1[0,2,12,0,8] =  1  &  W_1[0,2,12,8,0] =  1  \\
W_1[0,3,0,9,12] =  1  &  W_1[0,3,0,12,9] =  1  &  W_1[0,3,9,0,12] =  1  \\
W_1[0,3,9,12,0] =  1  &  W_1[0,3,12,0,9] =  1  &  W_1[0,3,12,9,0] =  1  \\
W_1[0,4,0,10,12] =  1  &  W_1[0,4,0,12,10] =  1  &  W_1[0,4,10,0,12] =  1  \\
W_1[0,4,10,12,0] =  1  &  W_1[0,4,12,0,10] =  1  &  W_1[0,4,12,10,0] =  1  \\
W_1[0,5,0,11,12] =  1  &  W_1[0,5,0,12,11] =  1  &  W_1[0,5,11,0,12] =  1  \\
W_1[0,5,11,12,0] =  1  &  W_1[0,5,12,0,11] =  1  &  W_1[0,5,12,11,0] =  1  \\
W_1[0,6,0,0,12] =  2  &  W_1[0,6,0,12,0] =  2  &  W_1[0,6,12,0,0] =  2  \\
W_1[0,7,0,1,12] =  1  &  W_1[0,7,0,12,1] =  1  &  W_1[0,7,1,0,12] =  1  \\
W_1[0,7,1,12,0] =  1  &  W_1[0,7,12,0,1] =  1  &  W_1[0,7,12,1,0] =  1  \\
W_1[0,8,0,2,12] =  1  &  W_1[0,8,0,12,2] =  1  &  W_1[0,8,2,0,12] =  1  \\
W_1[0,8,2,12,0] =  1  &  W_1[0,8,12,0,2] =  1  &  W_1[0,8,12,2,0] =  1  \\
W_1[0,9,0,3,12] =  1  &  W_1[0,9,0,12,3] =  1  &  W_1[0,9,3,0,12] =  1  \\
W_1[0,9,3,12,0] =  1  &  W_1[0,9,12,0,3] =  1  &  W_1[0,9,12,3,0] =  1  \\
W_1[0,10,0,4,12] =  1  &  W_1[0,10,0,12,4] =  1  &  W_1[0,10,4,0,12] =  1  \\
W_1[0,10,4,12,0] =  1  &  W_1[0,10,12,0,4] =  1  &  W_1[0,10,12,4,0] =  1  \\
W_1[0,11,0,5,12] =  1  &  W_1[0,11,0,12,5] =  1  &  W_1[0,11,5,0,12] =  1  \\
W_1[0,11,5,12,0] =  1  &  W_1[0,11,12,0,5] =  1  &  W_1[0,11,12,5,0] =  1  \\
W_1[0,12,0,0,6] =  2  &  W_1[0,12,0,1,7] =  1  &  W_1[0,12,0,2,8] =  1  \\
W_1[0,12,0,3,9] =  1  &  W_1[0,12,0,4,10] =  1  &  W_1[0,12,0,5,11] =  1  \\
W_1[0,12,0,6,0] =  2  &  W_1[0,12,0,7,1] =  1  &  W_1[0,12,0,8,2] =  1  \\
W_1[0,12,0,9,3] =  1  &  W_1[0,12,0,10,4] =  1  &  W_1[0,12,0,11,5] =  1  \\
W_1[0,12,1,0,7] =  1  &  W_1[0,12,1,7,0] =  1  &  W_1[0,12,2,0,8] =  1  \\
W_1[0,12,2,8,0] =  1  &  W_1[0,12,3,0,9] =  1  &  W_1[0,12,3,9,0] =  1  \\
W_1[0,12,4,0,10] =  1  &  W_1[0,12,4,10,0] =  1  &  W_1[0,12,5,0,11] =  1  \\
W_1[0,12,5,11,0] =  1  &  W_1[0,12,6,0,0] =  2  &  W_1[0,12,7,0,1] =  1  \\
W_1[0,12,7,1,0] =  1  &  W_1[0,12,8,0,2] =  1  &  W_1[0,12,8,2,0] =  1  \\
W_1[0,12,9,0,3] =  1  &  W_1[0,12,9,3,0] =  1  &  W_1[0,12,10,0,4] =  1  \\
W_1[0,12,10,4,0] =  1  &  W_1[0,12,11,0,5] =  1  &  W_1[0,12,11,5,0] =  1  \\
W_1[1,0,1,6,12] =  1  &  W_1[1,0,1,12,6] =  1  &  W_1[1,0,6,1,12] =  1  \\
W_1[1,0,6,12,1] =  1  &  W_1[1,0,12,1,6] =  1  &  W_1[1,0,12,6,1] =  1  \\
W_1[1,1,0,6,12] =  1  &  W_1[1,1,0,12,6] =  1  &  W_1[1,1,1,7,12] =  2  \\
W_1[1,1,1,12,7] =  2  &  W_1[1,1,2,8,12] =  1  &  W_1[1,1,2,12,8] =  1  \\
W_1[1,1,3,9,12] =  1  &  W_1[1,1,3,12,9] =  1  &  W_1[1,1,4,10,12] =  1  \\
W_1[1,1,4,12,10] =  1  &  W_1[1,1,5,11,12] =  1  &  W_1[1,1,5,12,11] =  1  \\
W_1[1,1,6,0,12] =  1  &  W_1[1,1,6,12,0] =  1  &  W_1[1,1,7,1,12] =  2  \\
W_1[1,1,7,12,1] =  2  &  W_1[1,1,8,2,12] =  1  &  W_1[1,1,8,12,2] =  1  \\
W_1[1,1,9,3,12] =  1  &  W_1[1,1,9,12,3] =  1  &  W_1[1,1,10,4,12] =  1  \\
W_1[1,1,10,12,4] =  1  &  W_1[1,1,11,5,12] =  1  &  W_1[1,1,11,12,5] =  1  \\
W_1[1,1,12,0,6] =  1  &  W_1[1,1,12,1,7] =  2  &  W_1[1,1,12,2,8] =  1  \\
W_1[1,1,12,3,9] =  1  &  W_1[1,1,12,4,10] =  1  &  W_1[1,1,12,5,11] =  1  \\
W_1[1,1,12,6,0] =  1  &  W_1[1,1,12,7,1] =  2  &  W_1[1,1,12,8,2] =  1  \\
W_1[1,1,12,9,3] =  1  &  W_1[1,1,12,10,4] =  1  &  W_1[1,1,12,11,5] =  1  \\
W_1[1,2,1,8,12] =  1  &  W_1[1,2,1,12,8] =  1  &  W_1[1,2,8,1,12] =  1  \\
W_1[1,2,8,12,1] =  1  &  W_1[1,2,12,1,8] =  1  &  W_1[1,2,12,8,1] =  1  \\
W_1[1,3,1,9,12] =  1  &  W_1[1,3,1,12,9] =  1  &  W_1[1,3,9,1,12] =  1  \\
W_1[1,3,9,12,1] =  1  &  W_1[1,3,12,1,9] =  1  &  W_1[1,3,12,9,1] =  1  \\
W_1[1,4,1,10,12] =  1  &  W_1[1,4,1,12,10] =  1  &  W_1[1,4,10,1,12] =  1  \\
W_1[1,4,10,12,1] =  1  &  W_1[1,4,12,1,10] =  1  &  W_1[1,4,12,10,1] =  1  \\
W_1[1,5,1,11,12] =  1  &  W_1[1,5,1,12,11] =  1  &  W_1[1,5,11,1,12] =  1  \\
W_1[1,5,11,12,1] =  1  &  W_1[1,5,12,1,11] =  1  &  W_1[1,5,12,11,1] =  1  \\
W_1[1,6,0,1,12] =  1  &  W_1[1,6,0,12,1] =  1  &  W_1[1,6,1,0,12] =  1  \\
W_1[1,6,1,12,0] =  1  &  W_1[1,6,12,0,1] =  1  &  W_1[1,6,12,1,0] =  1  \\
W_1[1,7,1,1,12] =  2  &  W_1[1,7,1,12,1] =  2  &  W_1[1,7,12,1,1] =  2  \\
W_1[1,8,1,2,12] =  1  &  W_1[1,8,1,12,2] =  1  &  W_1[1,8,2,1,12] =  1  \\
W_1[1,8,2,12,1] =  1  &  W_1[1,8,12,1,2] =  1  &  W_1[1,8,12,2,1] =  1  \\
W_1[1,9,1,3,12] =  1  &  W_1[1,9,1,12,3] =  1  &  W_1[1,9,3,1,12] =  1  \\
W_1[1,9,3,12,1] =  1  &  W_1[1,9,12,1,3] =  1  &  W_1[1,9,12,3,1] =  1  \\
W_1[1,10,1,4,12] =  1  &  W_1[1,10,1,12,4] =  1  &  W_1[1,10,4,1,12] =  1  \\
W_1[1,10,4,12,1] =  1  &  W_1[1,10,12,1,4] =  1  &  W_1[1,10,12,4,1] =  1  \\
W_1[1,11,1,5,12] =  1  &  W_1[1,11,1,12,5] =  1  &  W_1[1,11,5,1,12] =  1  \\
W_1[1,11,5,12,1] =  1  &  W_1[1,11,12,1,5] =  1  &  W_1[1,11,12,5,1] =  1  \\
W_1[1,12,0,1,6] =  1  &  W_1[1,12,0,6,1] =  1  &  W_1[1,12,1,0,6] =  1  \\
W_1[1,12,1,1,7] =  2  &  W_1[1,12,1,2,8] =  1  &  W_1[1,12,1,3,9] =  1  \\
W_1[1,12,1,4,10] =  1  &  W_1[1,12,1,5,11] =  1  &  W_1[1,12,1,6,0] =  1  \\
W_1[1,12,1,7,1] =  2  &  W_1[1,12,1,8,2] =  1  &  W_1[1,12,1,9,3] =  1  \\
W_1[1,12,1,10,4] =  1  &  W_1[1,12,1,11,5] =  1  &  W_1[1,12,2,1,8] =  1  \\
W_1[1,12,2,8,1] =  1  &  W_1[1,12,3,1,9] =  1  &  W_1[1,12,3,9,1] =  1  \\
W_1[1,12,4,1,10] =  1  &  W_1[1,12,4,10,1] =  1  &  W_1[1,12,5,1,11] =  1  \\
W_1[1,12,5,11,1] =  1  &  W_1[1,12,6,0,1] =  1  &  W_1[1,12,6,1,0] =  1  \\
W_1[1,12,7,1,1] =  2  &  W_1[1,12,8,1,2] =  1  &  W_1[1,12,8,2,1] =  1  \\
W_1[1,12,9,1,3] =  1  &  W_1[1,12,9,3,1] =  1  &  W_1[1,12,10,1,4] =  1  \\
W_1[1,12,10,4,1] =  1  &  W_1[1,12,11,1,5] =  1  &  W_1[1,12,11,5,1] =  1  \\
W_1[2,0,2,6,12] =  1  &  W_1[2,0,2,12,6] =  1  &  W_1[2,0,6,2,12] =  1  \\
W_1[2,0,6,12,2] =  1  &  W_1[2,0,12,2,6] =  1  &  W_1[2,0,12,6,2] =  1  \\
W_1[2,1,2,7,12] =  1  &  W_1[2,1,2,12,7] =  1  &  W_1[2,1,7,2,12] =  1  \\
W_1[2,1,7,12,2] =  1  &  W_1[2,1,12,2,7] =  1  &  W_1[2,1,12,7,2] =  1  \\
W_1[2,2,0,6,12] =  1  &  W_1[2,2,0,12,6] =  1  &  W_1[2,2,1,7,12] =  1  \\
W_1[2,2,1,12,7] =  1  &  W_1[2,2,2,8,12] =  2  &  W_1[2,2,2,12,8] =  2  \\
W_1[2,2,3,9,12] =  1  &  W_1[2,2,3,12,9] =  1  &  W_1[2,2,4,10,12] =  1  \\
W_1[2,2,4,12,10] =  1  &  W_1[2,2,5,11,12] =  1  &  W_1[2,2,5,12,11] =  1  \\
W_1[2,2,6,0,12] =  1  &  W_1[2,2,6,12,0] =  1  &  W_1[2,2,7,1,12] =  1  \\
W_1[2,2,7,12,1] =  1  &  W_1[2,2,8,2,12] =  2  &  W_1[2,2,8,12,2] =  2  \\
W_1[2,2,9,3,12] =  1  &  W_1[2,2,9,12,3] =  1  &  W_1[2,2,10,4,12] =  1  \\
W_1[2,2,10,12,4] =  1  &  W_1[2,2,11,5,12] =  1  &  W_1[2,2,11,12,5] =  1  \\
W_1[2,2,12,0,6] =  1  &  W_1[2,2,12,1,7] =  1  &  W_1[2,2,12,2,8] =  2  \\
W_1[2,2,12,3,9] =  1  &  W_1[2,2,12,4,10] =  1  &  W_1[2,2,12,5,11] =  1  \\
W_1[2,2,12,6,0] =  1  &  W_1[2,2,12,7,1] =  1  &  W_1[2,2,12,8,2] =  2  \\
W_1[2,2,12,9,3] =  1  &  W_1[2,2,12,10,4] =  1  &  W_1[2,2,12,11,5] =  1  \\
W_1[2,3,2,9,12] =  1  &  W_1[2,3,2,12,9] =  1  &  W_1[2,3,9,2,12] =  1  \\
W_1[2,3,9,12,2] =  1  &  W_1[2,3,12,2,9] =  1  &  W_1[2,3,12,9,2] =  1  \\
W_1[2,4,2,10,12] =  1  &  W_1[2,4,2,12,10] =  1  &  W_1[2,4,10,2,12] =  1  \\
W_1[2,4,10,12,2] =  1  &  W_1[2,4,12,2,10] =  1  &  W_1[2,4,12,10,2] =  1  \\
W_1[2,5,2,11,12] =  1  &  W_1[2,5,2,12,11] =  1  &  W_1[2,5,11,2,12] =  1  \\
W_1[2,5,11,12,2] =  1  &  W_1[2,5,12,2,11] =  1  &  W_1[2,5,12,11,2] =  1  \\
W_1[2,6,0,2,12] =  1  &  W_1[2,6,0,12,2] =  1  &  W_1[2,6,2,0,12] =  1  \\
W_1[2,6,2,12,0] =  1  &  W_1[2,6,12,0,2] =  1  &  W_1[2,6,12,2,0] =  1  \\
W_1[2,7,1,2,12] =  1  &  W_1[2,7,1,12,2] =  1  &  W_1[2,7,2,1,12] =  1  \\
W_1[2,7,2,12,1] =  1  &  W_1[2,7,12,1,2] =  1  &  W_1[2,7,12,2,1] =  1  \\
W_1[2,8,2,2,12] =  2  &  W_1[2,8,2,12,2] =  2  &  W_1[2,8,12,2,2] =  2  \\
W_1[2,9,2,3,12] =  1  &  W_1[2,9,2,12,3] =  1  &  W_1[2,9,3,2,12] =  1  \\
W_1[2,9,3,12,2] =  1  &  W_1[2,9,12,2,3] =  1  &  W_1[2,9,12,3,2] =  1  \\
W_1[2,10,2,4,12] =  1  &  W_1[2,10,2,12,4] =  1  &  W_1[2,10,4,2,12] =  1  \\
W_1[2,10,4,12,2] =  1  &  W_1[2,10,12,2,4] =  1  &  W_1[2,10,12,4,2] =  1  \\
W_1[2,11,2,5,12] =  1  &  W_1[2,11,2,12,5] =  1  &  W_1[2,11,5,2,12] =  1  \\
W_1[2,11,5,12,2] =  1  &  W_1[2,11,12,2,5] =  1  &  W_1[2,11,12,5,2] =  1  \\
W_1[2,12,0,2,6] =  1  &  W_1[2,12,0,6,2] =  1  &  W_1[2,12,1,2,7] =  1  \\
W_1[2,12,1,7,2] =  1  &  W_1[2,12,2,0,6] =  1  &  W_1[2,12,2,1,7] =  1  \\
W_1[2,12,2,2,8] =  2  &  W_1[2,12,2,3,9] =  1  &  W_1[2,12,2,4,10] =  1  \\
W_1[2,12,2,5,11] =  1  &  W_1[2,12,2,6,0] =  1  &  W_1[2,12,2,7,1] =  1  \\
W_1[2,12,2,8,2] =  2  &  W_1[2,12,2,9,3] =  1  &  W_1[2,12,2,10,4] =  1  \\
W_1[2,12,2,11,5] =  1  &  W_1[2,12,3,2,9] =  1  &  W_1[2,12,3,9,2] =  1  \\
W_1[2,12,4,2,10] =  1  &  W_1[2,12,4,10,2] =  1  &  W_1[2,12,5,2,11] =  1  \\
W_1[2,12,5,11,2] =  1  &  W_1[2,12,6,0,2] =  1  &  W_1[2,12,6,2,0] =  1  \\
W_1[2,12,7,1,2] =  1  &  W_1[2,12,7,2,1] =  1  &  W_1[2,12,8,2,2] =  2  \\
W_1[2,12,9,2,3] =  1  &  W_1[2,12,9,3,2] =  1  &  W_1[2,12,10,2,4] =  1  \\
W_1[2,12,10,4,2] =  1  &  W_1[2,12,11,2,5] =  1  &  W_1[2,12,11,5,2] =  1  \\
W_1[3,0,3,6,12] =  1  &  W_1[3,0,3,12,6] =  1  &  W_1[3,0,6,3,12] =  1  \\
W_1[3,0,6,12,3] =  1  &  W_1[3,0,12,3,6] =  1  &  W_1[3,0,12,6,3] =  1  \\
W_1[3,1,3,7,12] =  1  &  W_1[3,1,3,12,7] =  1  &  W_1[3,1,7,3,12] =  1  \\
W_1[3,1,7,12,3] =  1  &  W_1[3,1,12,3,7] =  1  &  W_1[3,1,12,7,3] =  1  \\
W_1[3,2,3,8,12] =  1  &  W_1[3,2,3,12,8] =  1  &  W_1[3,2,8,3,12] =  1  \\
W_1[3,2,8,12,3] =  1  &  W_1[3,2,12,3,8] =  1  &  W_1[3,2,12,8,3] =  1  \\
W_1[3,3,0,6,12] =  1  &  W_1[3,3,0,12,6] =  1  &  W_1[3,3,1,7,12] =  1  \\
W_1[3,3,1,12,7] =  1  &  W_1[3,3,2,8,12] =  1  &  W_1[3,3,2,12,8] =  1  \\
W_1[3,3,3,9,12] =  2  &  W_1[3,3,3,12,9] =  2  &  W_1[3,3,4,10,12] =  1  \\
W_1[3,3,4,12,10] =  1  &  W_1[3,3,5,11,12] =  1  &  W_1[3,3,5,12,11] =  1  \\
W_1[3,3,6,0,12] =  1  &  W_1[3,3,6,12,0] =  1  &  W_1[3,3,7,1,12] =  1  \\
W_1[3,3,7,12,1] =  1  &  W_1[3,3,8,2,12] =  1  &  W_1[3,3,8,12,2] =  1  \\
W_1[3,3,9,3,12] =  2  &  W_1[3,3,9,12,3] =  2  &  W_1[3,3,10,4,12] =  1  \\
W_1[3,3,10,12,4] =  1  &  W_1[3,3,11,5,12] =  1  &  W_1[3,3,11,12,5] =  1  \\
W_1[3,3,12,0,6] =  1  &  W_1[3,3,12,1,7] =  1  &  W_1[3,3,12,2,8] =  1  \\
W_1[3,3,12,3,9] =  2  &  W_1[3,3,12,4,10] =  1  &  W_1[3,3,12,5,11] =  1  \\
W_1[3,3,12,6,0] =  1  &  W_1[3,3,12,7,1] =  1  &  W_1[3,3,12,8,2] =  1  \\
W_1[3,3,12,9,3] =  2  &  W_1[3,3,12,10,4] =  1  &  W_1[3,3,12,11,5] =  1  \\
W_1[3,4,3,10,12] =  1  &  W_1[3,4,3,12,10] =  1  &  W_1[3,4,10,3,12] =  1  \\
W_1[3,4,10,12,3] =  1  &  W_1[3,4,12,3,10] =  1  &  W_1[3,4,12,10,3] =  1  \\
W_1[3,5,3,11,12] =  1  &  W_1[3,5,3,12,11] =  1  &  W_1[3,5,11,3,12] =  1  \\
W_1[3,5,11,12,3] =  1  &  W_1[3,5,12,3,11] =  1  &  W_1[3,5,12,11,3] =  1  \\
W_1[3,6,0,3,12] =  1  &  W_1[3,6,0,12,3] =  1  &  W_1[3,6,3,0,12] =  1  \\
W_1[3,6,3,12,0] =  1  &  W_1[3,6,12,0,3] =  1  &  W_1[3,6,12,3,0] =  1  \\
W_1[3,7,1,3,12] =  1  &  W_1[3,7,1,12,3] =  1  &  W_1[3,7,3,1,12] =  1  \\
W_1[3,7,3,12,1] =  1  &  W_1[3,7,12,1,3] =  1  &  W_1[3,7,12,3,1] =  1  \\
W_1[3,8,2,3,12] =  1  &  W_1[3,8,2,12,3] =  1  &  W_1[3,8,3,2,12] =  1  \\
W_1[3,8,3,12,2] =  1  &  W_1[3,8,12,2,3] =  1  &  W_1[3,8,12,3,2] =  1  \\
W_1[3,9,3,3,12] =  2  &  W_1[3,9,3,12,3] =  2  &  W_1[3,9,12,3,3] =  2  \\
W_1[3,10,3,4,12] =  1  &  W_1[3,10,3,12,4] =  1  &  W_1[3,10,4,3,12] =  1  \\
W_1[3,10,4,12,3] =  1  &  W_1[3,10,12,3,4] =  1  &  W_1[3,10,12,4,3] =  1  \\
W_1[3,11,3,5,12] =  1  &  W_1[3,11,3,12,5] =  1  &  W_1[3,11,5,3,12] =  1  \\
W_1[3,11,5,12,3] =  1  &  W_1[3,11,12,3,5] =  1  &  W_1[3,11,12,5,3] =  1  \\
W_1[3,12,0,3,6] =  1  &  W_1[3,12,0,6,3] =  1  &  W_1[3,12,1,3,7] =  1  \\
W_1[3,12,1,7,3] =  1  &  W_1[3,12,2,3,8] =  1  &  W_1[3,12,2,8,3] =  1  \\
W_1[3,12,3,0,6] =  1  &  W_1[3,12,3,1,7] =  1  &  W_1[3,12,3,2,8] =  1  \\
W_1[3,12,3,3,9] =  2  &  W_1[3,12,3,4,10] =  1  &  W_1[3,12,3,5,11] =  1  \\
W_1[3,12,3,6,0] =  1  &  W_1[3,12,3,7,1] =  1  &  W_1[3,12,3,8,2] =  1  \\
W_1[3,12,3,9,3] =  2  &  W_1[3,12,3,10,4] =  1  &  W_1[3,12,3,11,5] =  1  \\
W_1[3,12,4,3,10] =  1  &  W_1[3,12,4,10,3] =  1  &  W_1[3,12,5,3,11] =  1  \\
W_1[3,12,5,11,3] =  1  &  W_1[3,12,6,0,3] =  1  &  W_1[3,12,6,3,0] =  1  \\
W_1[3,12,7,1,3] =  1  &  W_1[3,12,7,3,1] =  1  &  W_1[3,12,8,2,3] =  1  \\
W_1[3,12,8,3,2] =  1  &  W_1[3,12,9,3,3] =  2  &  W_1[3,12,10,3,4] =  1  \\
W_1[3,12,10,4,3] =  1  &  W_1[3,12,11,3,5] =  1  &  W_1[3,12,11,5,3] =  1  \\
W_1[4,0,4,6,12] =  1  &  W_1[4,0,4,12,6] =  1  &  W_1[4,0,6,4,12] =  1  \\
W_1[4,0,6,12,4] =  1  &  W_1[4,0,12,4,6] =  1  &  W_1[4,0,12,6,4] =  1  \\
W_1[4,1,4,7,12] =  1  &  W_1[4,1,4,12,7] =  1  &  W_1[4,1,7,4,12] =  1  \\
W_1[4,1,7,12,4] =  1  &  W_1[4,1,12,4,7] =  1  &  W_1[4,1,12,7,4] =  1  \\
W_1[4,2,4,8,12] =  1  &  W_1[4,2,4,12,8] =  1  &  W_1[4,2,8,4,12] =  1  \\
W_1[4,2,8,12,4] =  1  &  W_1[4,2,12,4,8] =  1  &  W_1[4,2,12,8,4] =  1  \\
W_1[4,3,4,9,12] =  1  &  W_1[4,3,4,12,9] =  1  &  W_1[4,3,9,4,12] =  1  \\
W_1[4,3,9,12,4] =  1  &  W_1[4,3,12,4,9] =  1  &  W_1[4,3,12,9,4] =  1  \\
W_1[4,4,0,6,12] =  1  &  W_1[4,4,0,12,6] =  1  &  W_1[4,4,1,7,12] =  1  \\
W_1[4,4,1,12,7] =  1  &  W_1[4,4,2,8,12] =  1  &  W_1[4,4,2,12,8] =  1  \\
W_1[4,4,3,9,12] =  1  &  W_1[4,4,3,12,9] =  1  &  W_1[4,4,4,10,12] =  2  \\
W_1[4,4,4,12,10] =  2  &  W_1[4,4,5,11,12] =  1  &  W_1[4,4,5,12,11] =  1  \\
W_1[4,4,6,0,12] =  1  &  W_1[4,4,6,12,0] =  1  &  W_1[4,4,7,1,12] =  1  \\
W_1[4,4,7,12,1] =  1  &  W_1[4,4,8,2,12] =  1  &  W_1[4,4,8,12,2] =  1  \\
W_1[4,4,9,3,12] =  1  &  W_1[4,4,9,12,3] =  1  &  W_1[4,4,10,4,12] =  2  \\
W_1[4,4,10,12,4] =  2  &  W_1[4,4,11,5,12] =  1  &  W_1[4,4,11,12,5] =  1  \\
W_1[4,4,12,0,6] =  1  &  W_1[4,4,12,1,7] =  1  &  W_1[4,4,12,2,8] =  1  \\
W_1[4,4,12,3,9] =  1  &  W_1[4,4,12,4,10] =  2  &  W_1[4,4,12,5,11] =  1  \\
W_1[4,4,12,6,0] =  1  &  W_1[4,4,12,7,1] =  1  &  W_1[4,4,12,8,2] =  1  \\
W_1[4,4,12,9,3] =  1  &  W_1[4,4,12,10,4] =  2  &  W_1[4,4,12,11,5] =  1  \\
W_1[4,5,4,11,12] =  1  &  W_1[4,5,4,12,11] =  1  &  W_1[4,5,11,4,12] =  1  \\
W_1[4,5,11,12,4] =  1  &  W_1[4,5,12,4,11] =  1  &  W_1[4,5,12,11,4] =  1  \\
W_1[4,6,0,4,12] =  1  &  W_1[4,6,0,12,4] =  1  &  W_1[4,6,4,0,12] =  1  \\
W_1[4,6,4,12,0] =  1  &  W_1[4,6,12,0,4] =  1  &  W_1[4,6,12,4,0] =  1  \\
W_1[4,7,1,4,12] =  1  &  W_1[4,7,1,12,4] =  1  &  W_1[4,7,4,1,12] =  1  \\
W_1[4,7,4,12,1] =  1  &  W_1[4,7,12,1,4] =  1  &  W_1[4,7,12,4,1] =  1  \\
W_1[4,8,2,4,12] =  1  &  W_1[4,8,2,12,4] =  1  &  W_1[4,8,4,2,12] =  1  \\
W_1[4,8,4,12,2] =  1  &  W_1[4,8,12,2,4] =  1  &  W_1[4,8,12,4,2] =  1  \\
W_1[4,9,3,4,12] =  1  &  W_1[4,9,3,12,4] =  1  &  W_1[4,9,4,3,12] =  1  \\
W_1[4,9,4,12,3] =  1  &  W_1[4,9,12,3,4] =  1  &  W_1[4,9,12,4,3] =  1  \\
W_1[4,10,4,4,12] =  2  &  W_1[4,10,4,12,4] =  2  &  W_1[4,10,12,4,4] =  2  \\
W_1[4,11,4,5,12] =  1  &  W_1[4,11,4,12,5] =  1  &  W_1[4,11,5,4,12] =  1  \\
W_1[4,11,5,12,4] =  1  &  W_1[4,11,12,4,5] =  1  &  W_1[4,11,12,5,4] =  1  \\
W_1[4,12,0,4,6] =  1  &  W_1[4,12,0,6,4] =  1  &  W_1[4,12,1,4,7] =  1  \\
W_1[4,12,1,7,4] =  1  &  W_1[4,12,2,4,8] =  1  &  W_1[4,12,2,8,4] =  1  \\
W_1[4,12,3,4,9] =  1  &  W_1[4,12,3,9,4] =  1  &  W_1[4,12,4,0,6] =  1  \\
W_1[4,12,4,1,7] =  1  &  W_1[4,12,4,2,8] =  1  &  W_1[4,12,4,3,9] =  1  \\
W_1[4,12,4,4,10] =  2  &  W_1[4,12,4,5,11] =  1  &  W_1[4,12,4,6,0] =  1  \\
W_1[4,12,4,7,1] =  1  &  W_1[4,12,4,8,2] =  1  &  W_1[4,12,4,9,3] =  1  \\
W_1[4,12,4,10,4] =  2  &  W_1[4,12,4,11,5] =  1  &  W_1[4,12,5,4,11] =  1  \\
W_1[4,12,5,11,4] =  1  &  W_1[4,12,6,0,4] =  1  &  W_1[4,12,6,4,0] =  1  \\
W_1[4,12,7,1,4] =  1  &  W_1[4,12,7,4,1] =  1  &  W_1[4,12,8,2,4] =  1  \\
W_1[4,12,8,4,2] =  1  &  W_1[4,12,9,3,4] =  1  &  W_1[4,12,9,4,3] =  1  \\
W_1[4,12,10,4,4] =  2  &  W_1[4,12,11,4,5] =  1  &  W_1[4,12,11,5,4] =  1  \\
W_1[5,0,5,6,12] =  1  &  W_1[5,0,5,12,6] =  1  &  W_1[5,0,6,5,12] =  1  \\
W_1[5,0,6,12,5] =  1  &  W_1[5,0,12,5,6] =  1  &  W_1[5,0,12,6,5] =  1  \\
W_1[5,1,5,7,12] =  1  &  W_1[5,1,5,12,7] =  1  &  W_1[5,1,7,5,12] =  1  \\
W_1[5,1,7,12,5] =  1  &  W_1[5,1,12,5,7] =  1  &  W_1[5,1,12,7,5] =  1  \\
W_1[5,2,5,8,12] =  1  &  W_1[5,2,5,12,8] =  1  &  W_1[5,2,8,5,12] =  1  \\
W_1[5,2,8,12,5] =  1  &  W_1[5,2,12,5,8] =  1  &  W_1[5,2,12,8,5] =  1  \\
W_1[5,3,5,9,12] =  1  &  W_1[5,3,5,12,9] =  1  &  W_1[5,3,9,5,12] =  1  \\
W_1[5,3,9,12,5] =  1  &  W_1[5,3,12,5,9] =  1  &  W_1[5,3,12,9,5] =  1  \\
W_1[5,4,5,10,12] =  1  &  W_1[5,4,5,12,10] =  1  &  W_1[5,4,10,5,12] =  1  \\
W_1[5,4,10,12,5] =  1  &  W_1[5,4,12,5,10] =  1  &  W_1[5,4,12,10,5] =  1  \\
W_1[5,5,0,6,12] =  1  &  W_1[5,5,0,12,6] =  1  &  W_1[5,5,1,7,12] =  1  \\
W_1[5,5,1,12,7] =  1  &  W_1[5,5,2,8,12] =  1  &  W_1[5,5,2,12,8] =  1  \\
W_1[5,5,3,9,12] =  1  &  W_1[5,5,3,12,9] =  1  &  W_1[5,5,4,10,12] =  1  \\
W_1[5,5,4,12,10] =  1  &  W_1[5,5,5,11,12] =  2  &  W_1[5,5,5,12,11] =  2  \\
W_1[5,5,6,0,12] =  1  &  W_1[5,5,6,12,0] =  1  &  W_1[5,5,7,1,12] =  1  \\
W_1[5,5,7,12,1] =  1  &  W_1[5,5,8,2,12] =  1  &  W_1[5,5,8,12,2] =  1  \\
W_1[5,5,9,3,12] =  1  &  W_1[5,5,9,12,3] =  1  &  W_1[5,5,10,4,12] =  1  \\
W_1[5,5,10,12,4] =  1  &  W_1[5,5,11,5,12] =  2  &  W_1[5,5,11,12,5] =  2  \\
W_1[5,5,12,0,6] =  1  &  W_1[5,5,12,1,7] =  1  &  W_1[5,5,12,2,8] =  1  \\
W_1[5,5,12,3,9] =  1  &  W_1[5,5,12,4,10] =  1  &  W_1[5,5,12,5,11] =  2  \\
W_1[5,5,12,6,0] =  1  &  W_1[5,5,12,7,1] =  1  &  W_1[5,5,12,8,2] =  1  \\
W_1[5,5,12,9,3] =  1  &  W_1[5,5,12,10,4] =  1  &  W_1[5,5,12,11,5] =  2  \\
W_1[5,6,0,5,12] =  1  &  W_1[5,6,0,12,5] =  1  &  W_1[5,6,5,0,12] =  1  \\
W_1[5,6,5,12,0] =  1  &  W_1[5,6,12,0,5] =  1  &  W_1[5,6,12,5,0] =  1  \\
W_1[5,7,1,5,12] =  1  &  W_1[5,7,1,12,5] =  1  &  W_1[5,7,5,1,12] =  1  \\
W_1[5,7,5,12,1] =  1  &  W_1[5,7,12,1,5] =  1  &  W_1[5,7,12,5,1] =  1  \\
W_1[5,8,2,5,12] =  1  &  W_1[5,8,2,12,5] =  1  &  W_1[5,8,5,2,12] =  1  \\
W_1[5,8,5,12,2] =  1  &  W_1[5,8,12,2,5] =  1  &  W_1[5,8,12,5,2] =  1  \\
W_1[5,9,3,5,12] =  1  &  W_1[5,9,3,12,5] =  1  &  W_1[5,9,5,3,12] =  1  \\
W_1[5,9,5,12,3] =  1  &  W_1[5,9,12,3,5] =  1  &  W_1[5,9,12,5,3] =  1  \\
W_1[5,10,4,5,12] =  1  &  W_1[5,10,4,12,5] =  1  &  W_1[5,10,5,4,12] =  1  \\
W_1[5,10,5,12,4] =  1  &  W_1[5,10,12,4,5] =  1  &  W_1[5,10,12,5,4] =  1  \\
W_1[5,11,5,5,12] =  2  &  W_1[5,11,5,12,5] =  2  &  W_1[5,11,12,5,5] =  2  \\
W_1[5,12,0,5,6] =  1  &  W_1[5,12,0,6,5] =  1  &  W_1[5,12,1,5,7] =  1  \\
W_1[5,12,1,7,5] =  1  &  W_1[5,12,2,5,8] =  1  &  W_1[5,12,2,8,5] =  1  \\
W_1[5,12,3,5,9] =  1  &  W_1[5,12,3,9,5] =  1  &  W_1[5,12,4,5,10] =  1  \\
W_1[5,12,4,10,5] =  1  &  W_1[5,12,5,0,6] =  1  &  W_1[5,12,5,1,7] =  1  \\
W_1[5,12,5,2,8] =  1  &  W_1[5,12,5,3,9] =  1  &  W_1[5,12,5,4,10] =  1  \\
W_1[5,12,5,5,11] =  2  &  W_1[5,12,5,6,0] =  1  &  W_1[5,12,5,7,1] =  1  \\
W_1[5,12,5,8,2] =  1  &  W_1[5,12,5,9,3] =  1  &  W_1[5,12,5,10,4] =  1  \\
W_1[5,12,5,11,5] =  2  &  W_1[5,12,6,0,5] =  1  &  W_1[5,12,6,5,0] =  1  \\
W_1[5,12,7,1,5] =  1  &  W_1[5,12,7,5,1] =  1  &  W_1[5,12,8,2,5] =  1  \\
W_1[5,12,8,5,2] =  1  &  W_1[5,12,9,3,5] =  1  &  W_1[5,12,9,5,3] =  1  \\
W_1[5,12,10,4,5] =  1  &  W_1[5,12,10,5,4] =  1  &  W_1[5,12,11,5,5] =  2  \\
\hline \hline
\caption{Tensor $W_1$ (multiplied by $12\sqrt{7}$).}
\label{Table:W1}
\end{longtable}

%%%%%%%%%%%%%%%%%%%%%%%%%%%%%%%%%%%%%%%%%%%%%%%%%%%%%%

%bibliography using bibtex
%\bibliographystyle{apsrev4-1}
%\bibliographystyle{abbrv}
\bibliography{draft_hamilt_SU4}

%%%%%%%%%%%%%%%%%%%%%%%%%%%%%%%%%%%%%%%%%%%%%%%%%%%%%%

%%%%%%%%%%%%%%%%%%%%%%%%%%%%%%%%%%%%%%%%%%%%%%%%%%%%%%

\end{document}

%%%%%%%%%%%%%%%%%%%%%%%%%%%%%%%%%%%%%%%%%%%%%%%%%%%%%%